\documentclass[aps, pra, amsmath,amssymb,aps,floatfix, twocolumn, superscriptaddress]{revtex4-2}

\usepackage[utf8]{inputenc}
\usepackage{graphicx}
\usepackage{dcolumn}
\usepackage{bm}
\usepackage{color}
\usepackage[normalem]{ulem}

\def\spose#1{\hbox to 0pt{#1\hss}}
\def\lesssim{\mathrel{\spose{\lower 3pt\hbox{$\mathchar"218$}}
 \raise 2.0pt\hbox{$\mathchar"13C$}}}
\def\gtrsim{\mathrel{\spose{\lower 3pt\hbox{$\mathchar"218$}}
 \raise 2.0pt\hbox{$\mathchar"13E$}}}

\usepackage[dvipsnames]{xcolor}

\usepackage{hyperref}
\usepackage{xr} 
\begin{document}

\title{Geometrical entanglement and alignment regulate self-organization in active ring polymer suspensions}
\author{Juan Pablo Miranda}
\affiliation{ Dep. Est. de la Materia, F\'isica T\'ermica y Electr\'onica, Universidad Complutense de Madrid, 28040 Madrid, Spain}
\affiliation{GISC - Grupo Interdisciplinar de Sistemas Complejos 28040 Madrid, Spain}
\email{juanpami@ucm.es}

\author{Emanuele Locatelli}%
 \email{emanuele.locatelli@unipd.it}
\affiliation{Department of Physics and Astronomy,  University  of  Padova, 35131 Padova,  Italy}
\affiliation{INFN, Sezione di Padova, via Marzolo 8, I-35131 Padova, Italy}

 \author{Cristian Micheletti}
\affiliation{Scuola Internazionale Superiore di Studi Avanzati (SISSA), Via Bonomea 265, I-34136 Trieste, Italy}
\email{michelet@sissa.it}

 \author{Demian Levis}
\affiliation{Computing and Understanding Collective Action (CUCA) Lab, Condensed Matter
Physics Department, Universitat de Barcelona, Marti i Franquès 1, 08028 Barcelona,
Spain}
\affiliation{ University of Barcelona Institute of Complex Systems (UBICS), Martí i Franquès 1,E08028 Barcelona, Spain }
\email{levis@ub.edu}

 \author{Chantal Valeriani}
\affiliation{ Dep. Est. de la Materia, F\'isica T\'ermica y Electr\'onica, Universidad Complutense de Madrid, 28040 Madrid, Spain}
\affiliation{GISC - Grupo Interdisciplinar de Sistemas Complejos 28040 Madrid, Spain}
\email{cvaleriani@ucm.es}

\begin{abstract}
$\null$\\$\null$\\{\bf ABSTRACT}\\ 
We study the emerging self-organization in active ring suspensions, focusing on how the rings' orientational order and geometric entanglement vary with density and spatial confinement. To quantify entanglement, we introduce the wrapping number, a pairwise measure of ring interpenetration, while orientational order is characterized by the alignment of the normal vectors to the rings' osculating planes. Both wrapping number and alignment distinguish active from passive systems, and their combination aptly identifies the self-organized states that emerge with the onset of activity. Mutual-information analysis reveals a significant correlation between alignment and wrapping number across all considered active conditions. However, self-organization displays a non-monotonic dependence on the activity-induced entanglement. Specifically, moderate wrapping stabilizes contacts of neighboring aligned rings, while excessive entanglement disrupts alignment. We show that this competition arises because increasing entanglement interferes with the planar conformations required to form aligned stacks. Given the simplicity of this microscopic mechanism, analogous effects may occur more generally in polymer systems where the degree of entanglement is regulated by out-of-equilibrium effects.
\end{abstract}
\date{\today}
\maketitle

\section{Introduction}

Topological features  are ubiquitous in systems comprising isolated or interacting polymers and fibers, manifesting as knots, links, and other persistent forms of entanglement~\cite{meluzzi2010biophysics,fielden2017molecular,orlandini2021topological,tubiana2024topology}. These topological constraints typically reflect in fundamentally distinct metric\cite{metzler2002equilibrium,moore2004topologically,virnau2005knots,ercolini2007fractal,Tezuka_TopologicalSynthesis_JACS01,DehaghaniCatenaneExponent2020,LiDoubleScaling2021,D4SM00694A,D4SM00729H}, dynamical\cite{Rubinstein_Nature_2008,PPA_Science04,jain2017simulations,Soh_PRL19,RauscherDynamicsMelt2020,Rosa2020a,amici2021topological,ChiarantoniMichelettiRigidity2022,OConnor2022,mao2023diffusion,mao2024dynamics} and mechanical\cite{DoyleKnottedDNAPNAS,Caraglio_et_al_macromol_2017,farago2002pulling,ChiarantoniMichelettiNanochannel2023} properties compared to non-entangled systems. Various complex topologies have been identified in naturally-occurring molecules, both at equilibrium and in their biologically-functional states, from knotted proteins~\cite{virnau2006intricate,jackson2017fold, baiesi2017exploring,dabrowski2017topological} and isolated DNA molecules~\cite{rybenkov1993probability,jain2017simulations}, to genomic DNA\cite{valdes2018dna,valdes2019transcriptional,rosa2019topological}, including that of kinetoplasts which is organized in supramolecular DNA chainmails~\cite{chen1995topology,klotz2020equilibrium,he2023single,ramakrishnan_single_2024}.

Driving filamentous systems out-of-equilibrium can lead to intricate and persistent forms of entanglement, even when the nominal topological complexity is minimal or absent, such as in suspensions of initially unlinked and unknotted rings.
For example, long daisy chains of deadlocked ring polymers emerge in elongational flow \cite{o2020topological}, while scalar activity in ring melts, introduced by maintaining segments at different temperatures~\cite{chubak2020emergence,smrek2020active}, can generate a percolating network of deadlocks, resulting in a dynamical arrest of the system~\cite{micheletti2024topology}. Moreover, loop extrusion activity contributes to organizing genomic DNA {\em in vivo}~\cite{harju2025loop} and can modulate entanglement in models of polymer melts~\cite{conforto2024fluidification}. 
Finally, anomalous topological properties can emerge in polymer models driven out-of-equilibrium by vectorial activity, that is by a tangential (or polar) self-propulsion force parallel to the polymer backbone~\cite{isele2015self, bianco2018}.

In linear chains, tangential propulsion can suppress~\cite{foglino2019} or enhance~\cite{vatin2024conformation,li2024activity,vatin2025upsurge} the formation of knots, depending on the architecture of the chain: Using an active-passive diblock copolymer, knot formation can be tuned by regulating the formation and dissolution rates~\cite{vatin2025upsurge}. 
In dense suspensions of polar active polymers, one observes an enhancement of reptation~\cite{tejedor2019}, leading to peculiar effects such as cluster formation, connecting active polymers to active nematics \cite{oller2025computational}. At the microscopic scale, activity introduces characteristic length and time scales that regulate conformation and dynamics, contrasting the slow relaxation induced by entanglements in passive systems~\cite{ubertini2024universal}. 
At the macroscopic scale, activity crucially affects the rheological response, producing a giant increase of the stress plateau~\cite{breoni2025giant}. 
For ring polymers, polar activity gives rise to interesting phenomena both in two ~\cite{lamura2024excluded} and three dimensions~\cite{locatelli2021activity} at infinite dilution. In particular, in three dimensions, a sufficiently high self-propulsion induces swelling in short rings and collapse beyond a certain critical size. Instead, at finite density some of us have found that suspensions of short active rings display self-organized states at certain values of the monomer density and activity; notably, confinement {was found to} enhance the formation of such states~\cite{Miranda2023}.

In this work, we consider suspensions of short tangential active rings in bulk and under confinement, and characterize the self-organized states that emerge from the interplay of activity strength, density and confinement. As we show, the resulting combined action can either favour or contrast the planarity of the rings, leading to the formation or disruption of ordered stacks. 

The contrast between these emergent effects is studied by performing a network or clustering analysis based on suitable local observables. These include a nematic-like order parameter to detect ordered ring stacks, and a novel scalar observable, named wrapping number, that measures the complexity of the intertwining of unlinked rings. Importantly, we observe that significant intermingling hinders self-organization across all considered conditions, an effect that may hold more generally given that out-of-equilibrium driving often results in enhanced entanglement in polymer systems.  

\section{Model and Methods}
\subsection{Active ring polymer model}
We investigate the same active ring model as in Ref.~\cite{Miranda2023}. Briefly, we consider suspensions of active bead-spring ring polymers, composed by $N=76$ monomers each.  
Monomer pairs interact via a repulsive WCA potential \cite{WCA}
\begin{equation}
   U^{\mathrm{WCA}}(r)=\begin{cases}
4 \epsilon\left[\left(\sigma / r\right)^{12}-\left(\sigma / r\right)^{6}+\frac{1}{4}\right], \qquad &\text{$r < 2^{1 / 6}$} \sigma \\
0, &\text{otherwise,}
\end{cases}
\label{eq:wca}
\end{equation}
where $r$ is the distance between two monomers, $\sigma$ is their nominal diameter, $\epsilon = 50 k_B T$ and $k_B T$ is the thermal energy. We take $\sigma$ and $k_B T$ as the units of length and energy, respectively. 
Neighboring monomers along the contour of a ring are additionally subject to a FENE bonding potential
\begin{equation}
  U^{\mathrm{FENE}}(r)=\begin{cases}
-0.5 K R_{0}^{2} \ln \left[1-\left(r/ R_{0}\right)^{2}\right], & r \leq R_{0} \\
\infty, &\text{otherwise}
\end{cases}
\label{eq:fene}
\end{equation}
where $K=$30$\epsilon /\sigma^2$=1500$k_B T/\sigma^2$ and $R_0=1.5 \sigma$.
This set of values of the parameters $\sigma$, $\epsilon$, $K$, and $R_0$ avoids strand crossings and preserves at all times the unlinked and unknotted states of the rings set in the initial state~\cite{locatelli2021activity, Miranda2023}.

The activity is treated as self-propulsion force acting on each monomer, ${\bf f}^{\rm a}_{i}$. The active force has constant magnitude $f^a$ and, by construction, is parallel to the local tangent to the polymer backbone
\begin{equation}
{\bf f}^{\rm a}_{i} = f^a \frac{{\bf r}_{i+i} - {\bf r}_{i-i}}{|{\bf r}_{i+i} - {\bf r}_{i-i}|}.
\end{equation}

The strength of the activity is quantified by the P\'eclet number,
defined as $\mathrm{Pe}=f^a \sigma/k_B T$; we fix here $\mathrm{Pe}=10$. {As reported in~\cite{locatelli2021activity}, isolated active rings of length $N$ display swollen configurations at this value of $\mathrm{Pe}$; further, highly out-of-equilibrium self-organized states have also been reported at finite density~\cite{Miranda2023}.} 

We consider active ring suspensions in bulk and under lateral confinement; in the latter case, confinement is provided by two perfectly smooth, infinite planes placed at a distance $h$, {that is, at positions $z_w=0$ and $z_w=h$,} and orthogonal to the $z$ direction. The interaction between the flat walls and the monomers is purely repulsive and WCA-like (that is, Eq.~\eqref{eq:wca}), with {parameters} $\sigma_w=\sigma=1.$ and $\epsilon_w = k_B T$,
\begin{equation}
   U^{\mathrm{wall}}(z)=\begin{cases}
4 \epsilon_w \left[\left(\frac{\sigma_w}{z - z_w}  \right)^{12}-\left(\frac{\sigma_w}{z - z_w} \right)^{6}+\frac{1}{4}\right], \qquad &\text{$z < 2^{1 / 6}$} \sigma \\
0, &\text{otherwise.}
\end{cases}
\label{eq:wca_wall}
\end{equation}
We consider systems of $M$ active rings (with $M=500,1000$) characterised by different values of the monomer density $\rho= M \cdot N / V$, where $V$ is the volume of the simulation box: $\rho=$0.2, 0.3, 0.4, 0.5. In the confined case, we consider $h/\sigma=$9, 15, 21, 30, as in Ref.~\cite{Miranda2023}; we will refer to the bulk case as $h/\sigma=\infty$. For other simulation details, we refer to Ref.~\cite{Miranda2023}. We will consider long trajectories, sampling the steady state at low frequency (large time lags $\sim 10^7$ time steps or $\sim 10^3$ time units) to analyze the structure of the self-assembled states.

\begin{figure*}
    \centering
    \includegraphics[width=0.9\linewidth]{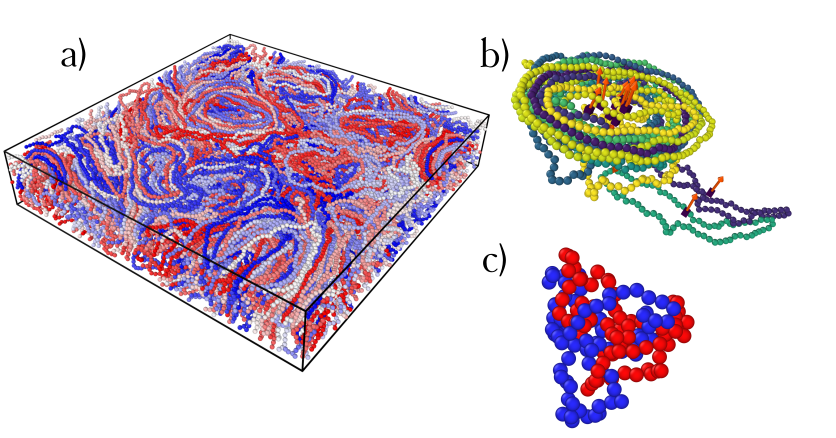}
       \caption{a) Snapshot of a confined system with density $\rho =0.3$ and $h/\sigma = 15$ where each ring polymer is highlighted by a different color. The top and bottom surfaces correspond to the confining walls.
    b) Snapshot of a cluster obtained from the alignment of the Voronoi neighbors {from the system shown in a)}. The purple cubes represent the positions of the centers of mass, and the orange arrow is the director of each ring (the magnitude of this director has been increased {in order to improve the readability of the figure}); a different color has been given to each polymer.
    c) Snapshot of a wrapped pair with {$W_{\mathrm{max}} \simeq 1.63$}.}
    \label{fig:intro}
\end{figure*}

\subsection{Quantifying ring entanglement: the wrapping number}\label{sec:wrap_def}

Since the rings remain unknotted and unlinked during dynamical evolution, we introduced a geometric observable to capture the entanglement between ring pairs, and specifically how much one ring wraps around another {(see Fig.~\ref{fig:intro}c))}.

The observable, which we name wrapping number, is defined as follows. For a ring pair{, denoted A and B}, we first consider all possible half-ring portions of ring A and compute the absolute value of their partial Gaussian linking integral with the entire ring B (technical details below). The largest value across all half-ring portions of ring A, $W_{A,B}$, is a nominal measure for how many times ring A encircles ring B. 

The roles of the two rings are then exchanged to compute $W_{B,A}$. The maximal wrapping number, $W_{\mathrm{max}}={\rm max}{(W_{A,B},W_{B,A})}$, is finally taken as the measure of entanglement for the ring pair.

From a technical point of view, we note the following. First, considering subportions of one of the rings is necessary because the Gaussian linking integral computed over the entire unlinked rings would be equal to zero.  Second, for computational efficiency, we first screen out ring pairs that are linearly separable, hence with $W_{\mathrm{max}}=0$. This geometrical test is performed with a perceptron scheme~\cite{krauth1987learning}, which requires a much smaller computational expenditure than calculating $W_{\mathrm{max}}$. We will refer to ring pairs that display a non vanishing wrapping ($W_{\mathrm{max}}>0$) as non linearly separable in the rest of the paper. Third, we compute the partial Gaussian linking integral with the signed crossings method~\cite{orlandini2021topological}. The procedure involves considering numerous projections (100 in our case) of the two oriented curves, namely, the piecewise linear backbones of the half-ring portion o and the entire partner ring. Next, the signs of the crossings in each projection are summed, and the projection-averaged sums yields the sought partial Gaussian linking integral.

For our analysis, we use the maximal wrapping number, $W_{\mathrm{max}}$, as follows. We consider two rings as significantly entangled, or significantly wrapped, if $W_{\mathrm{max}}\geq 0.5$, corresponding to at least half a turn around the partner ring. Consistent with this, we consider individual rings to be significantly entangled if they are significantly wrapped with one or more other rings.
Indicating with  $M_{\mathrm{wr}}$ the number of such rings, we use the ratio $\chi_{\mathrm{wr}} = M_{\mathrm{wr}}/M$, which lies in the [0:1] interval, to quantify the overall entanglement of the suspension.

\subsection{Metric and structural properties}
\label{sec:metric_method}

To characterize the emergence of  ordered supramolecular structures we used various metric observables, including the gyration tensor, defined as
\begin{equation}
    G_{\alpha \beta}=\frac{1}{N} \sum_{i=1}^{N}\left(\mathbf{r}_{i, \alpha}-\mathbf{r}_{\mathrm{cm}, \alpha}\right)\left(\mathbf{r}_{i, \beta}-\mathbf{r}_{\mathrm{cm}, \beta}\right) \ ,
\end{equation}
where the indices $\alpha$ and $\beta$ run over the three Cartesian coordinates ($x,y,z$) of the $N$ monomers of each ring, and
$\mathbf{r}_{\mathrm{cm}} = \frac{1}{N} \sum_{i=1}^{N} \mathbf{r}_{i}$ is the ring's center of mass.
For each ring, we compute the eigenvalues of the gyration tensor, $\lambda_1 \geq \lambda_2 \geq \lambda_3$, and define the \emph{director} of the ring, $\mathbf{d}$, as the eigenvector associated to the smallest eigenvalue, $\lambda_3$ 
\cite{alim2007shapes,catenanes2022Chiaratoni,stano2023cluster}. Thus, the director is the normal to the osculating plane of the ring.
When illustrating pairs or small groups of rings, we will represent the director as an arrow with origin in the centre of mass of the ring (see Fig.~\ref{fig:intro}b).\\

We use the rings' directors to compute the system-averaged nematic tensor
\begin{equation}
    T_{\alpha \beta}=\frac{1}{M} \sum_{i=1}^{M}\left(\frac{3}{2} \mathbf{d}_{i, \alpha} \mathbf{d}_{i, \beta} - \frac{1}{2} \delta_{\alpha  \beta}\right)
    \label{eq:nem1}
\end{equation}
where $\delta_{\alpha \beta}$ is the Kronecker delta symbol. The scalar nematic order parameter, $Q$ is then obtained by recasting Eq.~\eqref{eq:nem1} as,
\begin{equation}
    T_{\alpha \beta} = Q \left( \frac{3}{2} \mathbf{n}_{\alpha}  \mathbf{n}_{\beta} - \frac{1}{2} \delta_{\alpha \beta}\right)
\end{equation}
where $\mathbf{n}$ is the principal nematic director, through which one identifies $Q$ as the largest eigenvector of the nematic tensor.

\subsection{Clustering algorithm} \label{sec:MethodsCluster}

We use the local nematic alignment of ring directors to detect clusters of stacked rings, such as those visible in Fig.~\ref{fig:intro}b).
To this end, we take into account that two quasi-planar stacked rings, $i$ and $j$, are characterized by having $\left| \vec{d}_i \cdot \vec{d}_j \right| \approx 1$~\cite{stano2023cluster} as well as by the proximity of their centers of mass.

We establish proximity by identifying those pairs of rings that are nearest neighbours in a Voronoi space tesselation constructed from the positions of the centers of mass of all rings.
For each neighbouring ring pairs, we further test if their directors are significantly aligned, $\bigl|\mathbf{d}_i\cdot\mathbf{d}_j\bigr| > \delta_{\mathrm{cut}}$, with $\delta_{\mathrm{cut}}$ set to  $0.975$. 
The above information is then encoded in an alignment weighted adjacency matrix, $A$,
\begin{equation}
A_{ij} =
\begin{cases}
1, & \text{if $i$ and $j$ are neighbors and $\lvert \mathbf{d}_i \cdot \mathbf{d}_j \rvert > \delta_{\text{cut}}$,}\\
0, & \text{otherwise},
\end{cases}
\label{eq:Adjacency}
\end{equation}
which we used as input for the NetworkX~\cite{hagberg2008exploring} Python package to identify the clusters of stacked rings.

We note that this strategy, besides being physically transparent, is more precise at identifying stacked rings than clustering schemes using spatial proximity only. In addition, the joint criterion of proximity and directors' alignment is stringent enough that stacks can be reliably identified at all considered densities and activity levels without the necessity to resort to parametric clustering schemes such as DBSCAN, which nevertheless still represent a viable method for the systems at hand~\cite{Miranda2023}. 

\subsection{Mutual Information Analysis}
\label{sec:MI}
We assess the correlation between the degree of alignment of neighboring rings $\left| \vec{d}_i \cdot \vec{d}_j \right|$ and their wrapping number $W_{\mathrm{max}}$ via their mutual information ($\mathrm{MI}$) \cite{cover2012elements}. The MI measures how much information about one variable can be obtained by solely observing the other variable.
For two discrete variables, $X$ and $Y$, with possible outcomes $x_{1,...,n}$ and $y_{1,...,m}$ 
$\mathrm{MI}$ is defined as
\begin{equation}
\mathrm{MI}(X,Y)=\sum_{i=1}^n \sum_{j=1}^m P\left(x_i, y_j\right) \log \frac{P\left(x_i, y_j\right)}{P\left(x_i\right) P\left(y_j\right)},
\label{eq:MI}
\end{equation}
where $P\left(x_i, y_j\right)$ is the joint distribution, while $P\left(x_i\right)$ and $P\left(y_j\right)$
are the marginals. If $X$ and $Y$ are independent, then $\mathrm{MI}=0$.

For a given pair of {non linearly separable} rings,
we take $X$ and $Y$ to represent, respectively, the directors' alignment and the wrapping number, each binned in four intervals or levels. For $X=\left| \vec{d}_i \cdot \vec{d}_j \right|$, which ranges from 0 to 1, we take the following interval boundaries: $(0.0, 0.5, 0.75, 0.975, 1)$. Notice that the intervals have uneven length, to account for the fact that the distribution of $X$ is skewed towards 1. 
For $Y=W_{\mathrm{max}}$ we instead consider the following equispaced boundaries: {$(0, 0.375,0.75, 1.125, 1.5)$}. 

The $Y$ range accounts for the fact that wrapping numbers larger than 1.5 are very rarely observed, and such exceptional instances are assigned to the last interval. {We checked different possible choices for the interval boundaries of $X$ and $Y$: the values of the $\mathrm{MI}$ obtained were always compatible.}
To measure the MI correlation of the observables for nearby rings, we consider {$\sim 10^2$} samples of the system and for each of them we randomly extract a small number of ring pairs, picked among those that are nearest neighbors in the Voronoi tesselation.
To avoid introducing spurious correlations, {this set will contain each ring only once and its size, for each sample, is smaller than $M$.}
Overall, the MI is typically computed over {$\sim 10^4$} data points.\\

To assess the statistical significance of the observed MI, we compare it against a synthetic null distribution of MI values obtained by reshuffling the values of the $X$ variable across the pairs. The reshuffling obliterates the correlation in the joint probability distribution, while preserving the two marginal ones.
Next, we compute the average,
$\mathrm{MI}_{\rm null}$, 
and the standard deviation, $\sigma_{\mathrm{MI}_{\rm null}}$ for the mutual information of the null case. Finally, we compute a standard $Z$-score of the MI observed in the actual data, $\mathrm{MI}_{\mathrm{Obs}}$, by computing by how many standard deviations it exceeds the average value of the null model:
\begin{equation}
    \mathrm{Z-score}= \frac{\left| \mathrm{MI}_{\mathrm{Obs}}  - \mathrm{MI}_{\mathrm{null}}\right|}{\sigma_{\mathrm{MI}_{\rm null}}},
\end{equation}

Accordingly, the higher the Z-score the more significant is the observed correlation of the variables.

\section{Results}
\subsection{Entanglement analysis}

To characterize the pairwise ring entanglement, we analyze the maximal wrapping number for varying $\rho$ and $h$, with and without activity. As discussed in the Methods section , the maximal wrapping number quantifies of how much a ring encircles the partner one, and is a convenient geometrical measure of entanglement considering that, topologically, all rings are unlinked~\cite{micheletti2024topology}.

\begin{figure}[h!]
    \centering
    \includegraphics[width=0.45\textwidth]{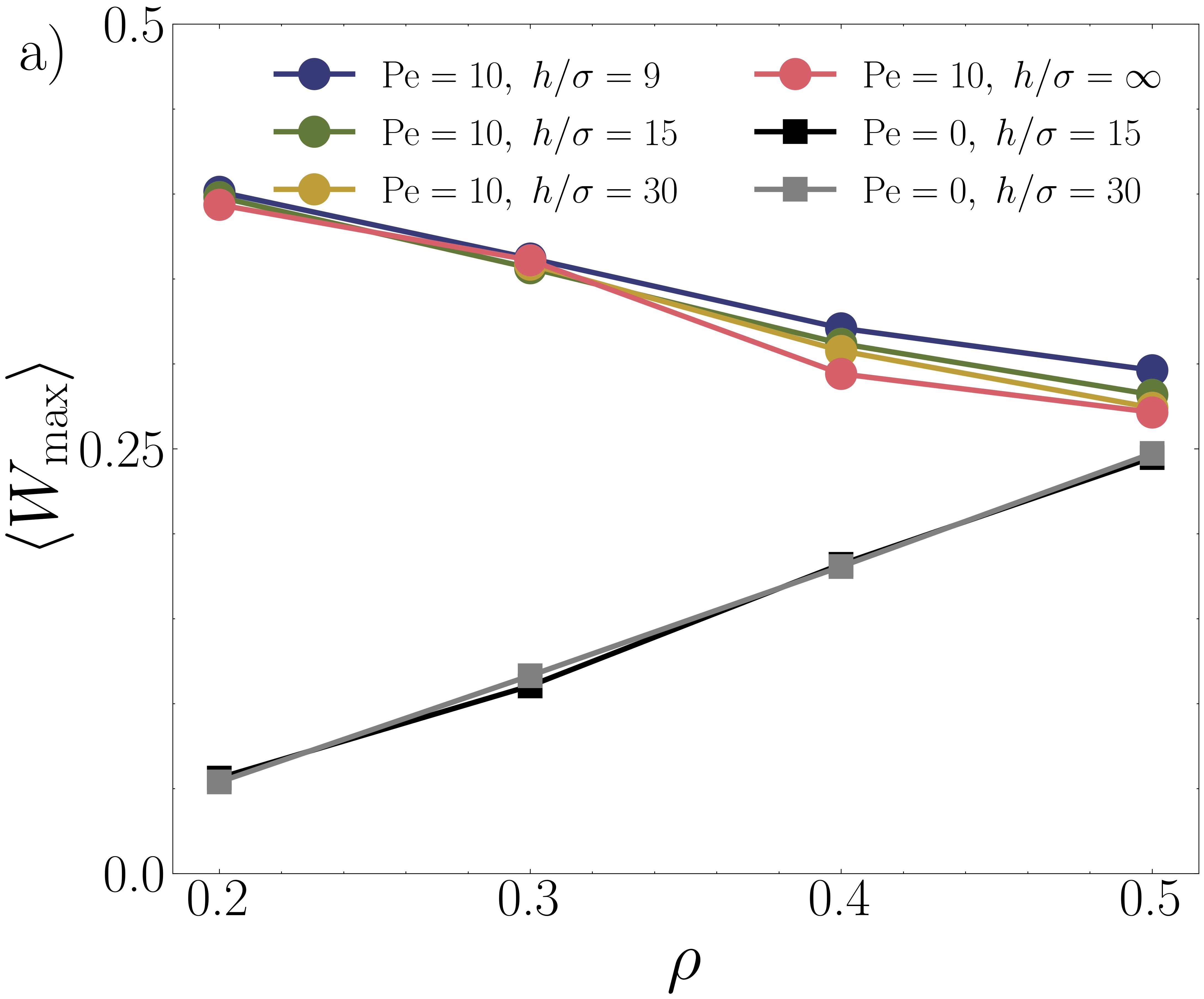}
    \includegraphics[width=0.45\textwidth]{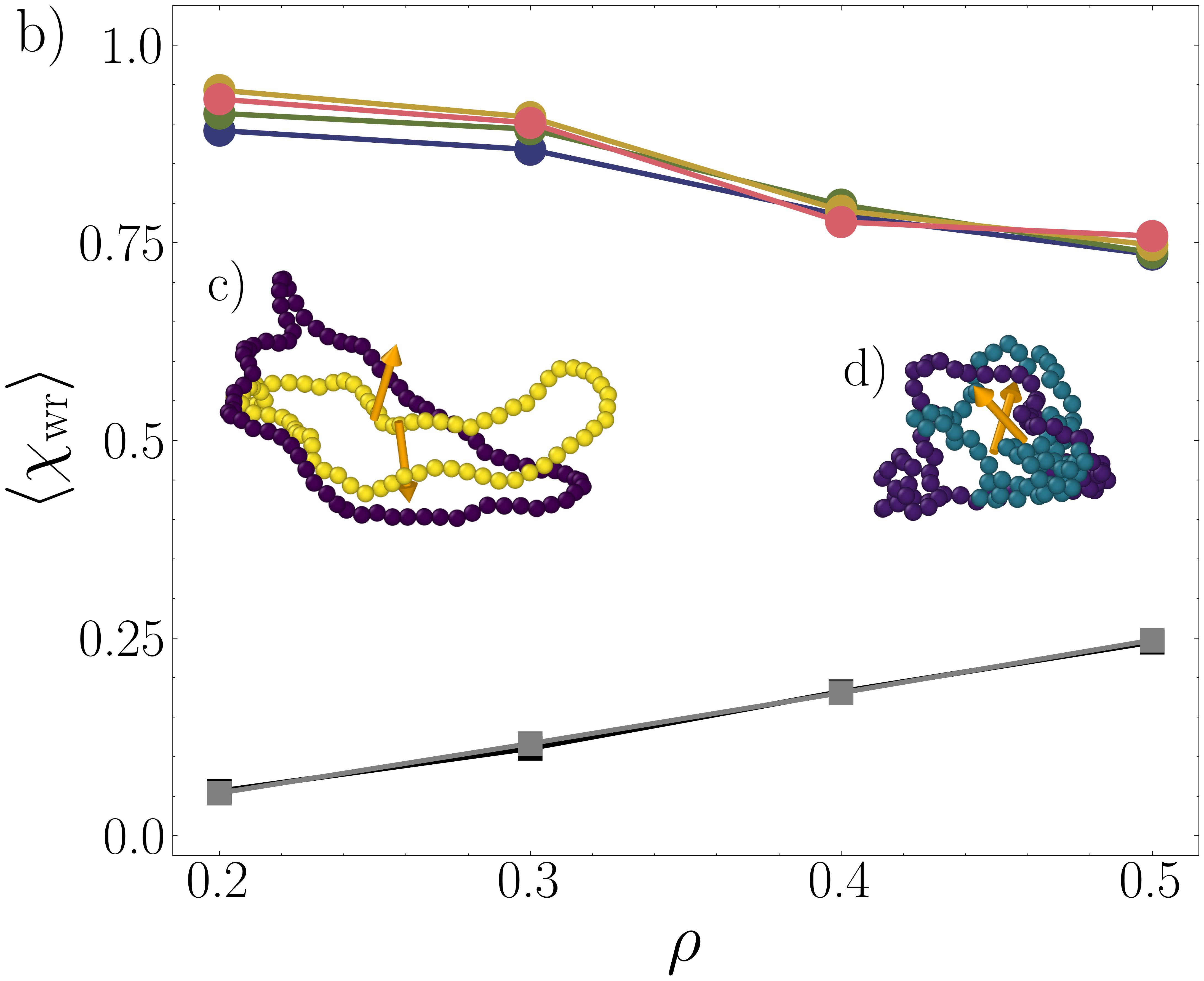}
  \caption{a) Average maximum wrapping, $\langle W_{\mathrm{max}} \rangle$, and b) Average fraction of significantly entangled {rings}  $\langle \chi_{\mathrm{wr}}\rangle$ (having $ W_{\mathrm{max}} \geq 0.5$ with at least another ring), as a function of the density $\rho$.
  In both figures, full dots represent active systems with $\mathrm{Pe} = 10$ at different confinement distances; 
  full squares represent passive systems ($\mathrm{Pe} = 0$) at different confinement distances.
   {Snapshots of wrapped pairs c) $\rho = 0.2 ,\ h/\sigma=21$, d) $\rho = 0.5 ,\ h/\sigma=21$; the arrows indicate the directors of the rings.}
  }
    \label{fig:TotalWrapping}
\end{figure}

As a first step, we consider the time-average of the maximum wrapping number, $\langle W_{\mathrm{max}} \rangle$, and of the fraction of {significantly} wrapped rings $\langle  \chi_{\mathrm{wr}} \rangle$, introduced in 
 Sec.~\ref{sec:wrap_def}. The time averages are taken at steady state. We remark that wrapping is defined for pairs of rings; thus, individual rings may possess different wrapping values. However, we find desirable to consider average features of individual rings, which should be less affected by the increase of crowding that comes from increasing density. 

As discussed in in Sec.~\ref{sec:wrap_def}, when computing $\chi_{\mathrm{wr}}$ we consider a ring to be significantly wrapped if $W_{\rm max} > 0.5$. 

Fig.~\ref{fig:TotalWrapping} shows these observables as a function of density $\rho$ for active systems ($\mathrm{Pe}=10$) at various confinements $h$, along with data for the passive counterparts ($\mathrm{Pe}=0$).

We first discuss the dependence of $\langle W_{\mathrm{max}} \rangle$ on $\rho$ (see Fig.~\ref{fig:TotalWrapping}(a)). We observe that $\langle W_{\mathrm{max}} \rangle$ is systematically larger in active systems than in passive ones and, in both cases, it has little dependence on confinement. Notably, $\langle W_{\mathrm{max}} \rangle$ decreases with increasing density $\rho$ in active systems, while the opposite is true in passive ones. This suggests that the wrapping number is a relevant observable for discriminating active and passive systems.

Indeed, a similar distinction emerges for $\langle  \chi_{\mathrm{wr}} \rangle$, reported in Fig.~\ref{fig:TotalWrapping}(b) as a function of $\rho$. Notice that, at low values of $\rho$, nearly all rings are significantly entangled with at least another ring; in contrast, this practically never happens in passive systems. 
As $\rho$ increases, the gap between the curves for active and passive systems reduces, though it sill persists at the largest $\rho$ considered.\\
The observed increase of $\langle W_{\mathrm{max}} \rangle$ with $\rho$ for passive systems is intuitively explained by the reduced separation of the  rings' centers of mass, and hence the increased overlap of the rings'  volumes, $R_g^{3/2}$\cite{kremer1990dynamics,brown1998computer}. 
Instead, in active systems, the rings organize in clusters and assume an oblate,  disk-like shape. The latter is affected by the contact interaction with nearby rings, which alter the local active forces and induce ring collapse\cite{locatelli2021activity}. 

Thus, with increasing density, the fraction of active rings that are collapsed, and hence not significantly involved in entanglements, increases.

It is also interesting to investigate the spatial organization of significantly entangled rings, especially in the presence of confinement.
To this end, we calculated the normalized number of wrapped rings $\chi_{\mathrm{wr}}(z)$ along the confinement direction, $z$, measured relative to the midpoint, $z/\sigma=h/2$. Specifically, we divided the system into slabs parallel to the $xy$ plane, and calculated $M_{\mathrm{wr}}(z)$ and $M(z)$ considering only rings with centers of mass within the slab. 

For comparison, we perform an analogous analysis for the bulk (unconfined) case, by averaging over slab subdivisions taken over all three Cartesian axes.

\begin{figure}[h!]
    \centering
     \includegraphics[width=0.45\textwidth]{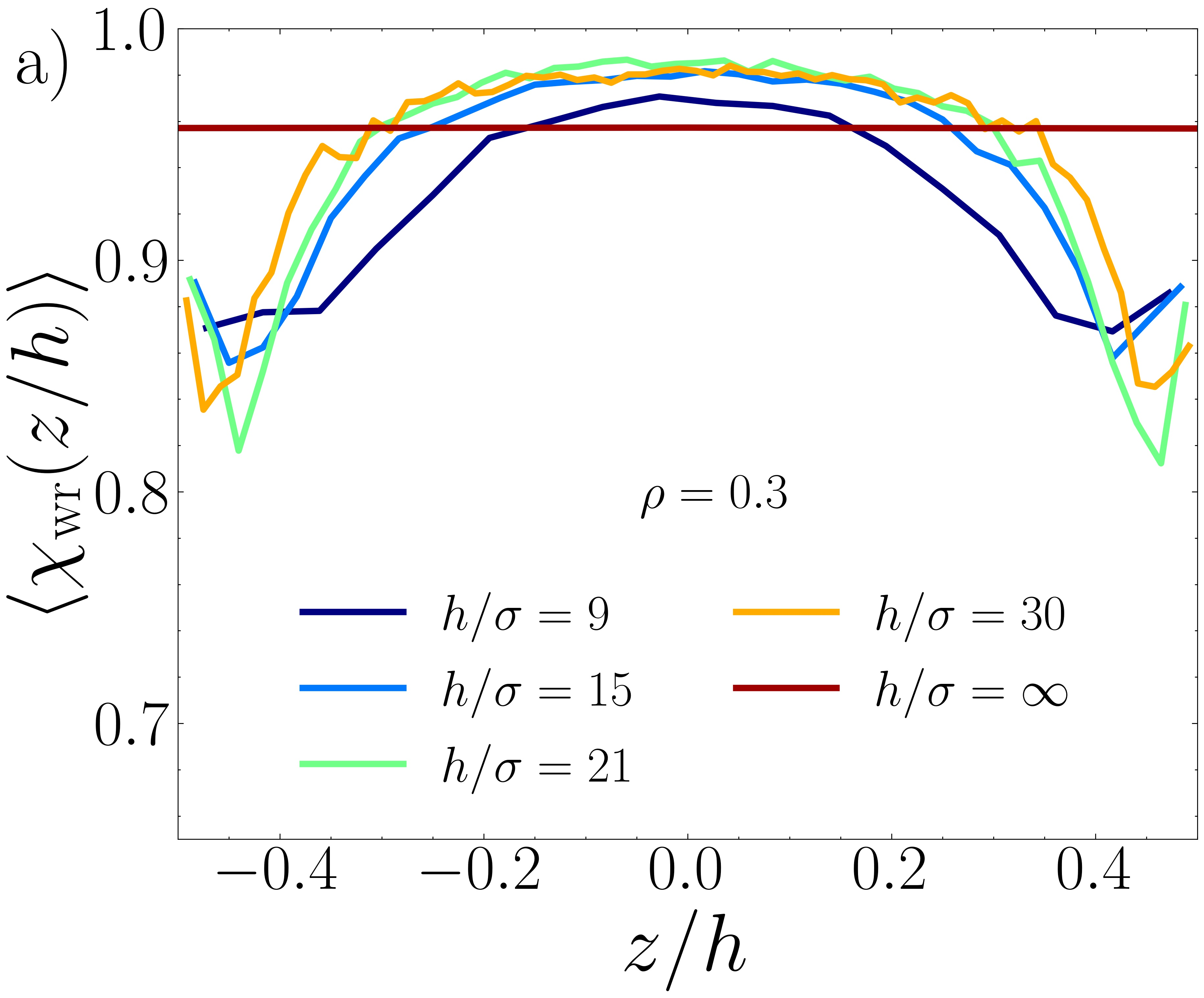}\\
     \includegraphics[width=0.45\textwidth]{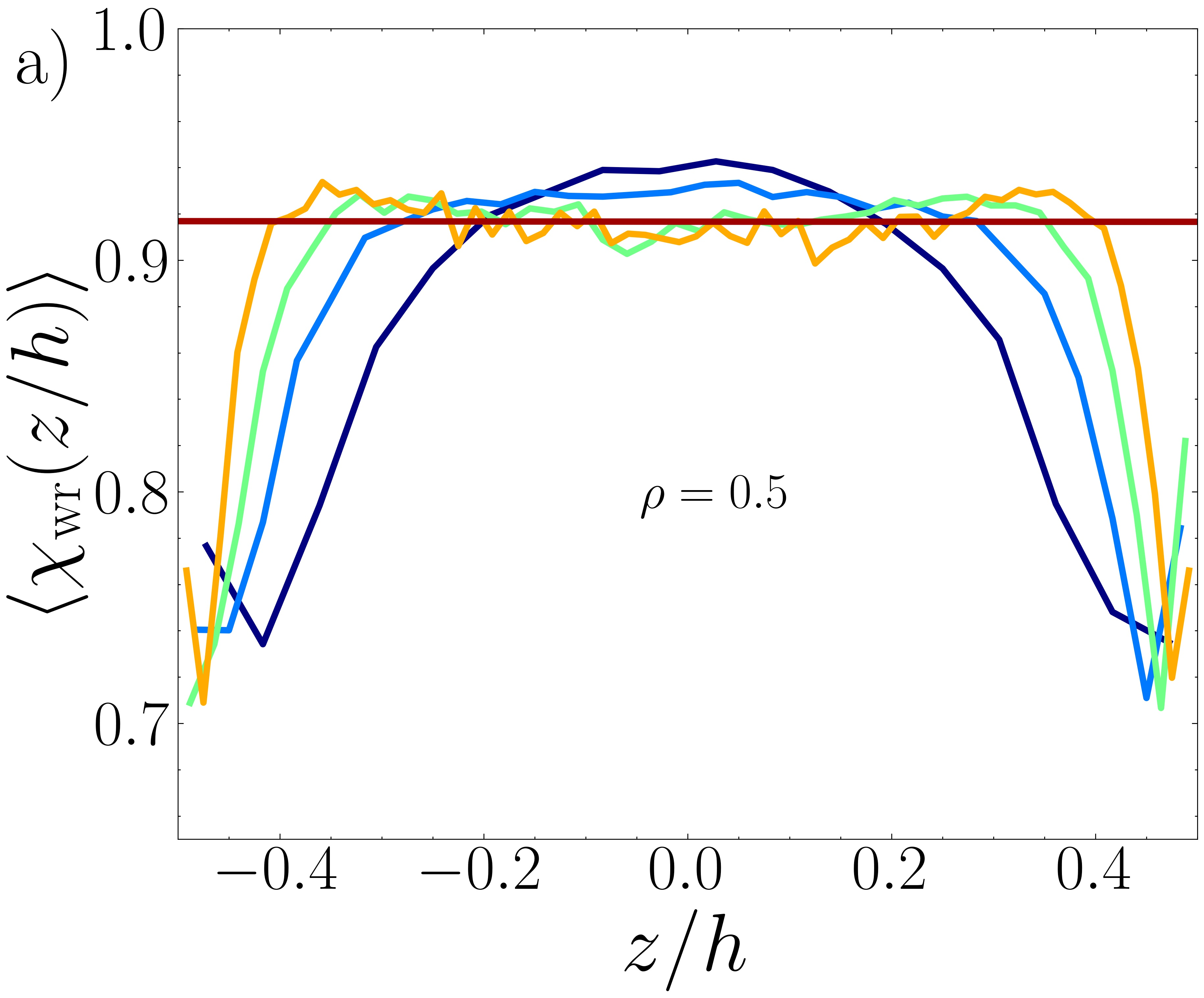}
  \caption{Average of the fraction of significant wrapped rings $\left\langle \chi_{\mathrm{wr}}\right\rangle$  as a function of $z/h$ for $h/\sigma = 9,15,21,30,\infty$ and a) $\rho=0.3$, b) $\rho=0.5$. }
\label{fig:chi_vs_z}
\end{figure}

We report the results in Fig.~\ref{fig:chi_vs_z} {(with additional data in the Supplemental Information)} at $\rho=0.2$ (Fig.~\ref{fig:chi_vs_z}(a)) and $\rho=0.5$ (Fig.~\ref{fig:chi_vs_z}(b)) across various levels of confinement, $h/\sigma$. We observe that both confinement and density reshape the spatial distribution of ring entanglements. Increasing $h$, the entanglement are distributed more uniformly across the confining planes, while strong confinement produces depletion layers at the bounding walls where rings are less likely to be significantly entangled.
 
The layers, which are approximately $1.5\sigma$-thick, are present at all densities, but become more sharply defined with increasing $\rho$. Furthermore, as the average fraction of significantly entangled ring pairs is nearly independent of $h$ (see Fig.~\ref{fig:TotalWrapping}), the presence of the depletion layers is compensated by an excess entanglement in the central region, compared to the bulk. This compensation effect diminishes with increasing $h/\sigma$ even at the largest considered density; interestingly, a noticeable difference between confined and bulk cases persists at $\rho=0.3$ for all the values of $h/\sigma$ considered. 
 
\subsection{Nematic order}

In the presence of activity, the rings self-organize into clusters of stacked rings \cite{Miranda2023}, as shown in Figure \ref{fig:intro}. As previously done for polyelectrolyte ring solutions~\cite{stano2023cluster}, we capture the orientational order of stacked rings by evaluating the alignment of the rings' directors and using it to define a global nematic order parameter, $\langle Q \rangle$, see Section \ref{sec:metric_method}.

\begin{figure}[b!]
    \centering
    \includegraphics[width=0.45\textwidth]{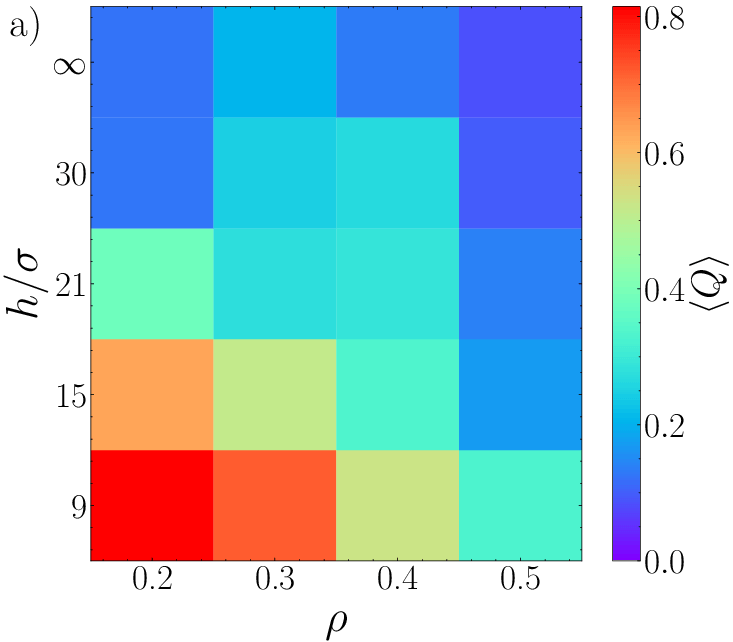}
    \includegraphics[width=0.45\textwidth]{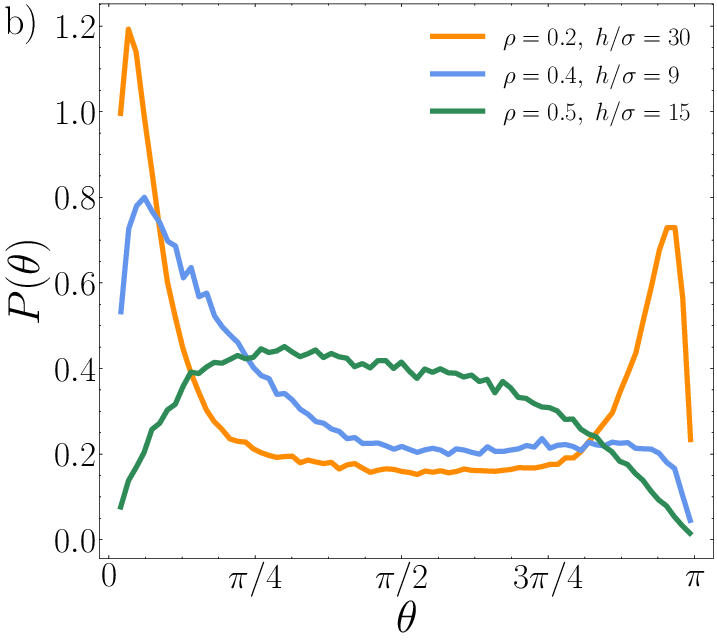}
    \includegraphics[width=0.49\textwidth]{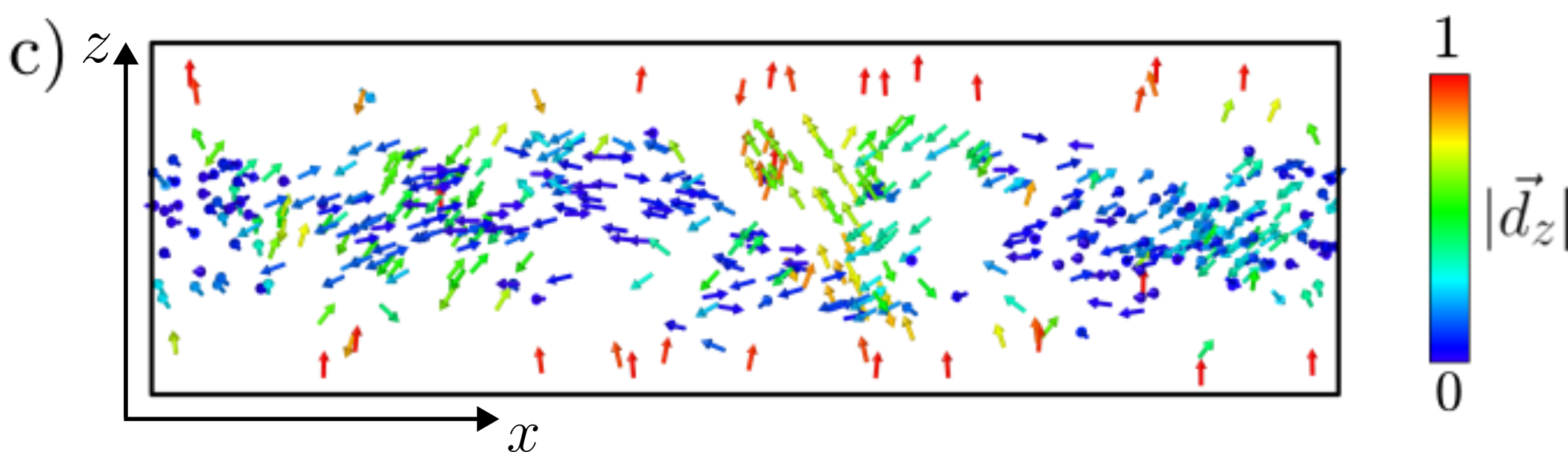}
    \caption{a) Color plot of the nematic order parameter $\langle Q \rangle$ (defined in Sec.~\ref{sec:metric_method}) in the $\rho$-$h/\sigma$ plane.
    b) Probability distribution of the angles between the directors of neighbouring rings, $P(\theta)$ for different values of $\rho$ and $h/\sigma$; neighbours are identified via a Voronoi tessellation (see Sec.~\ref{sec:MethodsCluster}). {c) Snapshot of the directors of the rings in the system with $\rho=0.3,\ h/\sigma = 21$; the color corresponds to the  absolute value of the $z$ component of each ring's director.}}     
    \label{fig:NematicOrder}
\end{figure}

Fig.~\ref{fig:NematicOrder} a) presents a heatmap of the nematic order parameter for various combinations of densities $\rho$ and confinement levels, $h/\sigma$.
Notice that $Q$ is maximal in strongly confined and low-density systems, and diminishes about monotonically for increasing $\rho$ and $h/\sigma$.

The effects are rationalized as follows. In strong confinement, stacking mostly occurs along the $z$ direction. This results in the stacks being parallel. As a consequence, rings are nematically ordered not only within stacks but also across them. As confinement is reduced, stacks become more randomly oriented and nematic order is progressively lost at the global level, persisting only within stacks. 
In addition, increasing $\rho$ reduces the planarity of contacting rings, thus hindering stack formation and lowering the nematic order within the stacks.

To further characterize the local alignment we considered the angle $\theta$ between the directors of pairs of neighboring rings, identified via a Voronoi tessellation (see Sec.~\ref {sec:MethodsCluster}).

Figure~\ref{fig:NematicOrder} b) shows the distribution $P(\theta)$ for systems at selected values $\rho$ and $h/\sigma$,  representative of the various degrees self-organized order induced by activity.

For the most ordered case (orange curve, $\rho=0.2$ and $h/\sigma=30$), the distribution features two peaks at $\theta= 0, \pi$, indicating strong nematic alignment between neighbors. 
In the intermediate case (blue curve, $\rho=0.4$ and $h/\sigma=9$), the peak at $\pi$ valushes while the one at 0 broadens, indicating that nematic alignment becomes unfavorable at these conditions. Finally, in the disordered regime (green curve, $\rho=0.5$ and $h/\sigma=15$), also the peak at $\theta = 0$ disappears, and the distribution broadens significantly. Here, crowding prevents disk-like ring conformations,  thus suppressing stacking and nematic order.

Fig.\ref{fig:NematicOrder} c) shows a cross ($xy$) projection of the nematic directors of the rings, $\vec{d}$, for for ($\rho=0.3,\ h/\sigma = 21$). The directors are color-coded according to the absolute value of their $z$-component, $\vec{d}_z$. The projected pattern shows a mild nematic order at the global level. Instead, the local nematic order is clearly strong.
Rings near the walls have directors approximately orthogonal to the walls, while in the center of the slab, local order between groups of rings is visible.
These observations motivate us to quantify and study 
the emergence of collective behavior through alignment between neighboring rings. 

\subsection{Alignment-based classification of the self-organization}

We identified bundles of stacked rings using the clustering procedure detailed in Sec.~\ref{sec:MethodsCluster}, which is based on the the nematic alignment of neighboring rings in the Voronoi tesselation.

To characterize the emergence of clusters, we used two main observables: the average number of clusters (stacks), $N_c$, and the cluster fraction $X_c$, that is, the fraction of rings that are involved in stack. An additional observable, the cluster size distribution, is provided in the Supplemental Information. 
\begin{figure}[h!]
    \centering
    \includegraphics[width=0.45\textwidth]{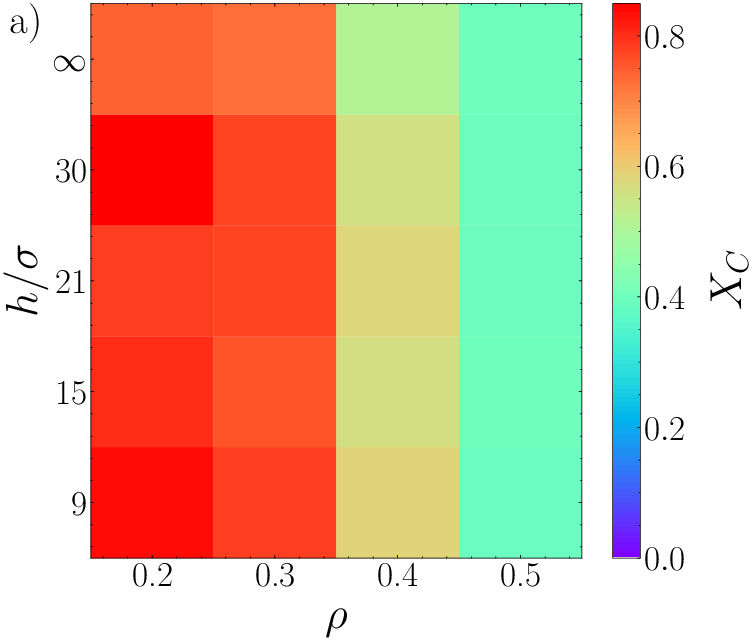}
    \includegraphics[width=0.45\textwidth]{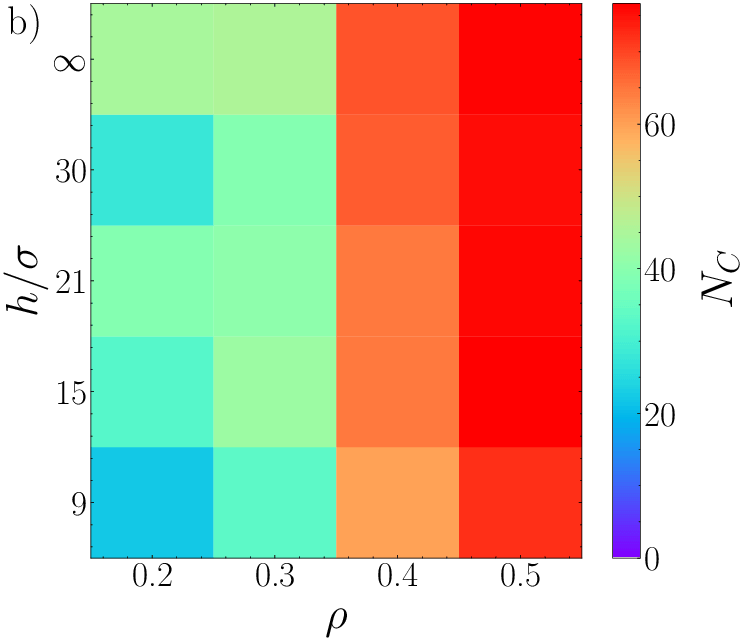}
    \caption{ Color plots of a) cluster fraction $X_C$  and  b) number of clusters $N_c$ in the $\rho$-$h/\sigma$ plane, obtained with the clustering method described in Sec~\ref{sec:MethodsCluster}.    }
    \label{fig:clustersize_vs_rho}
\end{figure}

The two main observables are shown in Fig.~\ref{fig:clustersize_vs_rho}, and are consistent with previous results~\cite{Miranda2023}. 
In general, $X_c$ decreases with increasing $\rho$, again due to crowding suppressing  ring planarity, and thus hindering stacking and alignment.
At low density, most rings belong to few large clusters, while at high density they form many small clusters (see Supplemental Information Fig. S5). For instance, at $\rho=0.5$, the largest clusters encompass $30-60$ rings, while at $\rho=0.2$ clusters with $100$ or more rings are typical. 

The absence of large clusters signals the loss of self-organization: at high density, small groups of aligned rings persist but, being dispersed, cannot merge into larger aggregates.

Interestingly, the alignment-based clustering suggests that confinement has a more limited ordering effect compared to density. 

At $\rho=0.2$, both  $X_c$ and $N_c$,  have a non-monotonic dependence on $h/\sigma$, as reflected by the color gradient in Fig.~\ref{fig:clustersize_vs_rho}. This effect weakens appreciably at higher density.

\begin{figure*}[t]
    \centering
    \includegraphics[width=0.85\textwidth]{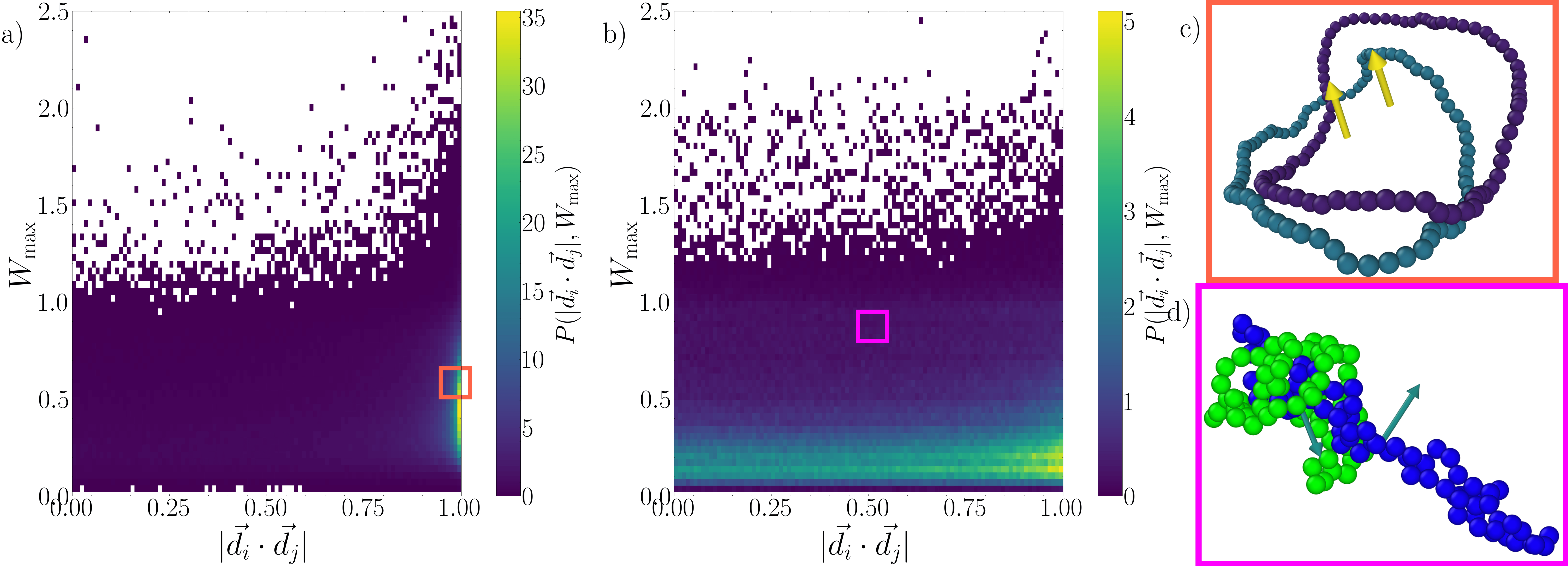}
    \caption{Joint probability distribution of the maximum wrapping and the alignment between pairs of non linearly separable Voronoi neighbors at $h/\sigma=21$  and a)  $\rho=0.2$; c) $\rho=0.5$.
    Panels c) and d) are snapshots corresponding to configurations of two wrapped rings from a) and b), respectively, {highlighted with a square matching the frame color of c) and d) panels.} Rings in c) are characterized by $W_{\mathrm{max}}\simeq 0.6$ and $\left| \vec{d}_i \cdot \vec{d}_j \right| \simeq 0.99$.
    Rings in d) are characterized by
    $W_{\mathrm{max}}\simeq 0.83$ and $\left| \vec{d}_i \cdot \vec{d}_j \right| \simeq 0.01$.
    The arrows in the snapshots represent the directors of the polymers and are located on the corresponding centers of mass. Rings are colored differently to help visualization, while the color of the arrows follows the same palette as the probability distribution in panels a),c), {yellow and purple corresponding to high and low alignment, respectively}.
 }
    \label{fig:joint_probability}
\end{figure*}

\subsection{Relation between {entanglement} and alignment}
The results presented so far indicate that activity induces various degrees of entanglement and alignment of the rings depending on the system density and confinement. Activity can introduce significant entanglement in neighboring rings compared to the passive case, and both alignment and entanglement decline as ring aggregation or stacking decreases.

To further explore the connection between ring entanglement and alignment of Voronoi-neighboring rings, we considered the joint probability distribution of their nematic order parameter and maximum wrapping number, $P(| \vec{d}_i \cdot \vec{d}_j |, W_{\mathrm{max}} )$.\\
Fig.~\ref{fig:joint_probability} presents $P(| \vec{d}_i \cdot \vec{d}_j |, W_{\mathrm{max}} )$ at fixed confinement $h/\sigma=21$ for two different densities $\rho=0.2,0.5$,  representative of strong and weak self-organization, respectively.

At $\rho=0.5$, the joint probability is almost independent of the nematic alignment at fixed $W_{\mathrm{max}}$, with only a slight enhancement near $| \vec{d}_i \cdot \vec{d}_j |=1$, indicating an overall lack of order and weak correlation of the two observables (Fig.~\ref{fig:joint_probability} b). The highest probability density occurs in a band at low wrapping values, approximately $0.1 \le W_{\mathrm{max}} \le 0.4$.

This picture changes radically for $\rho=0.2$, when self-organization is evident (Fig.~\ref{fig:joint_probability}a). The joint probability distribution $P(| \vec{d}_i \cdot \vec{d}_j | ,W_{\mathrm{max}} )$ features a sharp maximum at high nematic alignment and wrapping values in the $0.25 \le W_{\mathrm{max}} \le 0.75$ range, which include non-negligble though still partial ring entanglement. Both conditions are conducive to stacking, which is indeed significant at this density. 

Indeed, full ring entanglement, $W_{\mathrm{max}} \gtrsim 1$ or larger disfavors stacking because it is incompatible with ring planarity and limits ring interactions to a single partner ring. Conversely, rings involved in partially entangled pairs, $0.25 \le W_{\mathrm{max}} \le 0.75$, can form columnar stacks. 

The point is illustrated by the two selected configurations of Fig.~\ref{fig:joint_probability} c), d). The ring pair in the panel c), representing the self-organized state at $\rho=$ 0.2 and $h/\sigma=$ 21, is highly aligned ($| \vec{d}_i \cdot \vec{d}_j | \approx 0.99$) and has intermediate wrapping number ($W_{\mathrm{max}}\simeq 0.6$). In contrast, the pair in panel d), representing the disordered system at $\rho=$ 0.5 and $h/\sigma=$ 21, has a larger wrapping number ($W_{\mathrm{max}}\simeq 0.8$) and near-zero nematic alignment.\\

\begin{figure}[h!]
    \centering
    \includegraphics[width=0.45\textwidth]{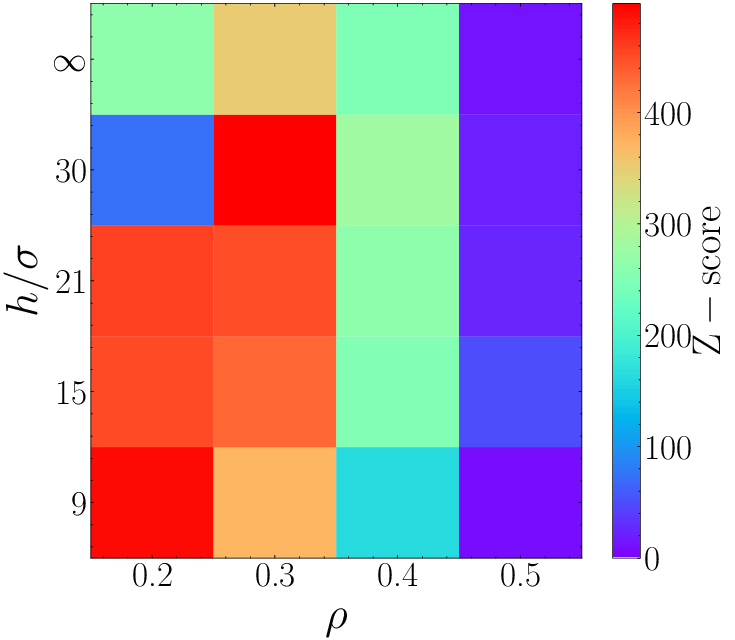}
    \caption{{Z-score test between alignment and wrapping of {non separable} Voronoi neighboring pairs in the $\rho$-$h/\sigma$ plane.}  }
    \label{fig:zscore_active}
\end{figure}

For a more quantitative connection between {entanglement} and alignment, we computed their Mutual Information and established the corresponding statistical significance via a Z-score analysis (see Sec.~\ref{sec:MI}).
The results of the analysis are reported in Fig.~\ref{fig:zscore_active}. For all systems with $\rho < 0.5$ the Z-score is much larger than unity, indicating that the observed MI exceeds by far the MI value expected in samples of the same size, but where the two variables are uncorrelated, see Sec.~\ref{sec:MI}. This confirms the significance of the correlation of entanglement and alignment in all systems where self-organization is observed ($\rho<0.5$). Conversely, across all  $h/\sigma$ values at $\rho = 0.5$, the Z-score values are compatible with those measured in passive systems (see Supplemental Information Fig.8). 

Thus, the statistical analysis supports the conclusion that wrapping number and alignment are statistically correlated in the emerging self-organized states, establishing a connection between geometric and topological properties for active ring suspensions. 
Importantly, this correlation is nonlinear. While for moderate activity, both ring alignment and entanglement (wrapping) increase,  high activity induces excessive ring wrapping which disrupts alignment. As a result, intermediate levels of activities are the most conducive for self-organization.

To further support this conclusion, we provide in III of the Supplemental Information the fraction of Voronoi neighboring rings that are both wrapped and aligned. This observable follows the same trend as the Z-score: it decreases with both increasing density and confinement. This is consistent with the behavior of the cluster fraction and reinforces the connection between entanglement, alignment, and self-organization.

Notably, albeit aligned pairs are a small fraction of the Voronoi neighbors, they suffice to seed self-organization. The fact that most aligned neighbors are also wrappeed further supports the idea that activity organizes the rings via an interplay of their planarity and compenetration.
\\

This connection provides a mechanistic basis for rationalizing the emergence of order and self-organization.
As already noted, the disk-like shape is necessary for stack formation, similarly to polyelectrolyte rings solutions. However, differently from these latter systems, no long range repulsion is present and thus there is no mechanism preventing entanglement. At the same time, cluster formation in charged rings occurs at very high density, while in active systems stacking is significant even at relatively low values of $\rho$. This suggests that some mechanism maintains rings in proximity, at least transiently. The analysis carried out in this work suggests that entanglement plays this role: however, the nature of the system is such that only specific values of the wrapping number are selected. 
Indeed, too much entanglement can be detrimental,  as activity-induced collisions lead to ring collapse, as observed in isolated long active rings~\cite{locatelli2021activity}.
In the systems considered, the ring length and its preferred shape intuitively favor shallow entanglements, since it is possible to partially overlap two rings at no energetic cost and without inducing collisions. However, if the entanglement is too little, rings will not remain in spatial proximity long enough for clustering to occur. 
These considerations help explain the observed ranges for the wrapping number that are conducive to stack formation, providing a rationale for the role of entanglements in the observed self-organization.

\section{Conclusions}
We have reported an extensive analysis of local geometrical features in suspensions of tangentially active short rings, both in bulk and under confinement, aiming to elucidate the emergence of self-organized states. On the one hand, we examined the geometrical entanglement through the wrapping number, which measures the degree of compenetration between rings. On the other hand, motivated by studies on polyelectrolyte ring suspensions, we looked at the alignment between the rings' orientation vectors, known as directors.\\ 
We found that the selected geometrical features can be used to discriminate between active and passive systems, as both entanglement and alignment are enhanced by tangential activity, with respect to the passive case. For example, the maximum wrapping as well as the fraction of significantly wrapped rings show a monotonically decreasing trend in active systems with increasing density; while the opposite trend is observed in the passive case. Further, the alignment probability between pairs of active rings is strongly affected by density: at low density, neighboring rings are predominantly aligned and, at high confinement, the system displays a relatively large nematic order parameter; the opposite happens at high density, the directors are rarely aligned.

To connect these trends to self-organization, we used a clustering algorithm based on the degree of alignment of neighboring rings. The resulting clusters are consistent with the previous characterization, supporting the  role of ring alignment as a key defining feature of the self-organized states.

To elucidate the connection between alignement and entanglement, we analyzed the joint probability distribution of wrapping number and nematic order parameter. The distribution differs significantly between systems with and without self-organization. In the former case, the probability concentrates in specific ranges of wrapping number and nematic order. In the latter, the nematic order parameter becomes more broadly distributed. To quantify this relationship, we resorted to the Mutual Information, and established its significance using the Z-score. This statistical analysis helped us establish a non-monotonic interdependence of the wrapping number and nematic order.

Based on these results, we conclude that the interplay of activity with density or confiment can endow polymer ring suspensions with qualitative properties not otherwise observed in passive systems. On the one hand, active rings can feature entanglements similar to dense polymer systems and, on the other they display orientational order similar to liquid crystalline materials.  It is this activity-induced dual nature that drives the swelling of the rings and enhances local alignment and entanglement of ring pairs.

While, to the best of our knowledge, suspensions of active ring polymers cannot be realized at the moment, these systems present an intriguing venue for a hybrid system, whose liquid-crystalline feature could be controlled to form, for example, nano-porous structures using an external field.


\vspace{2cm}
\textbf{Acknowledgments}
This study was funded in part by the European Union - NextGenerationEU, in the framework of the PRIN Project "The Physics of Chromosome Folding" (code: 2022R8YXMR, CUP: G53D23000820006) and by PNRR Mission 4, Component 2, Investment 1.4\_CN\_00000013\_CN-HPC: National Centre for HPC, Big Data and Quantum Computing - spoke 7 (CUP: G93C22000600001). D.L. acknowledges MCIU/AEI and DURSI for financial support under Projects No. PID2022-140407NB-C22 and 2021SGR-673, respectively. 
 C.V. acknowledges fundings
IHRC22/00002 and PID2022-140407NB-C21 from MINECO.

The views and opinions expressed are solely those of the authors and do not necessarily reflect those of the European Union, nor can the European Union be held responsible for them.

\vspace{1cm}
\textbf{Supporting Information} Additional results for the wrapping, the clustering method and the interplay between alignment are entanglement are provided in the Supporting Information.


\bibliography{bibliography}

\end{document}


\preprint{APS/123-QED}

\title{Geometrical entanglement and alignment regulate self-organization of active polymer rings suspensions - Supporting Information}

\author{Juan Pablo Miranda}
\affiliation{ Dep. Est. de la Materia, F\'isica T\'ermica y Electr\'onica, Universidad Complutense de Madrid, 28040 Madrid, Spain}
\affiliation{GISC - Grupo Interdisciplinar de Sistemas Complejos 28040 Madrid, Spain }

\author{Emanuele Locatelli}%
 \email{emanuele.locatelli@unipd.it}
\affiliation{Department of Physics and Astronomy,  University  of  Padova, 35131 Padova,  Italy}
\altaffiliation{INFN, Sezione di Padova, via Marzolo 8, I-35131 Padova, Italy}

 \author{Cristian Micheletti}
\affiliation{Scuola Internazionale Superiore di Studi Avanzati (SISSA),Via Bonomea 265, I-34136 Trieste, Italy}
\email{michelet@sissa.it}

 \author{Demian Levis}
\affiliation{Computing and Understanding Collective Action (CUCA) Lab, Condensed Matter
Physics Department, Universitat de Barcelona, Marti i Franquès 1, 08028 Barcelona,
Spain}
\affiliation{ University of Barcelona Institute of Complex Systems (UBICS), Martí i Franquès 1,E08028 Barcelona, Spain }
\email{levis@ub.edu}

 \author{Chantal Valeriani}
\affiliation{ Dep. Est. de la Materia, F\'isica T\'ermica y Electr\'onica, Universidad Complutense de Madrid, 28040 Madrid, Spain}
\affiliation{GISC - Grupo Interdisciplinar de Sistemas Complejos 28040 Madrid, Spain
}
\email{cvaleriani@ucm.es}

\maketitle

\section{Additional results on wrapping}

\subsection{Probability distribution of max wrapping}

We briefly report the probability distribution of the maximum wrapping, defined in the main text, comparing active and passive systems at the same confinement conditions.
\begin{figure}
    \centering
     \includegraphics[width=0.5\textwidth]{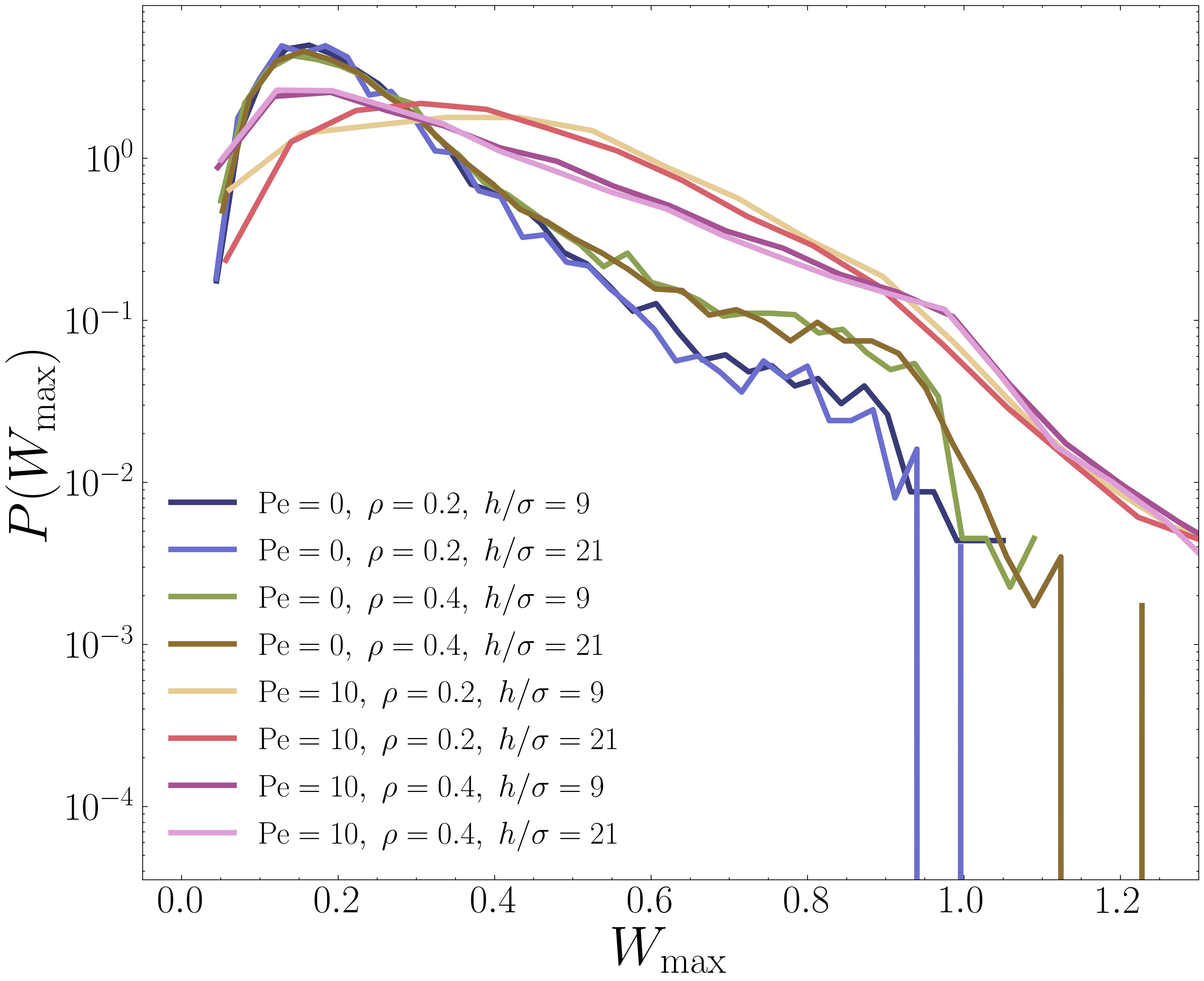}
  \caption{Probability distribution of the max wrapping $W_\mathrm{max}$ for systems of rings, active ($\mathrm{Pe}=10$) or passive ($\mathrm{Pe}=0$) at different values of the monomer density $\rho$ and of the separation $h/\sigma$ between the confining walls. 
  }
\label{fig:maxwrap_distro}    
\end{figure}
We report these results in Fig.~\ref{fig:maxwrap_distro}. We first focus on passive systems ($\mathrm{Pe}=0$): the distribution of the wrapping maintains the same shape and remains peaked around the same value upon changing $\rho$ and $h/\sigma$.In additiion, note that the most probable value is quite low ($W_{\mathrm{max}}\simeq 0.2$, leading to the low average value reported in the main text. In contrast, the distribution of the maximum wrapping depends both on the density and on the degree of confinement in active systems. At $\rho=0.2$, the most probable value is higher than in the passive case ($W_{\mathrm{max}}\simeq 0.4$) and the distribution has a longer tail, meaning that ring pairs with larger wrapping are much more probable. With increasing $\rho$, the most probable value decreases and becomes compatible with its passive counterpart; however, the distribution remains more fat tailed, suggesting a more preeminent role of the wrapping in active systems.  

\subsection{Spatial organization of wrapped rings}

We report here complementary data on the spatial organization of wrapped rings, as defined in the main text. Briefly, we take the number or rings that are significantly wrapped ($W > 0.5$) with at least one other ring at different values of the height $z$ from the centre of the box, normalized by the number of rings at the same height; we name this quantity $\chi_{\mathrm{wp}}(z)$.
\begin{figure*}[b]
    \centering
     \includegraphics[width=0.32\textwidth]{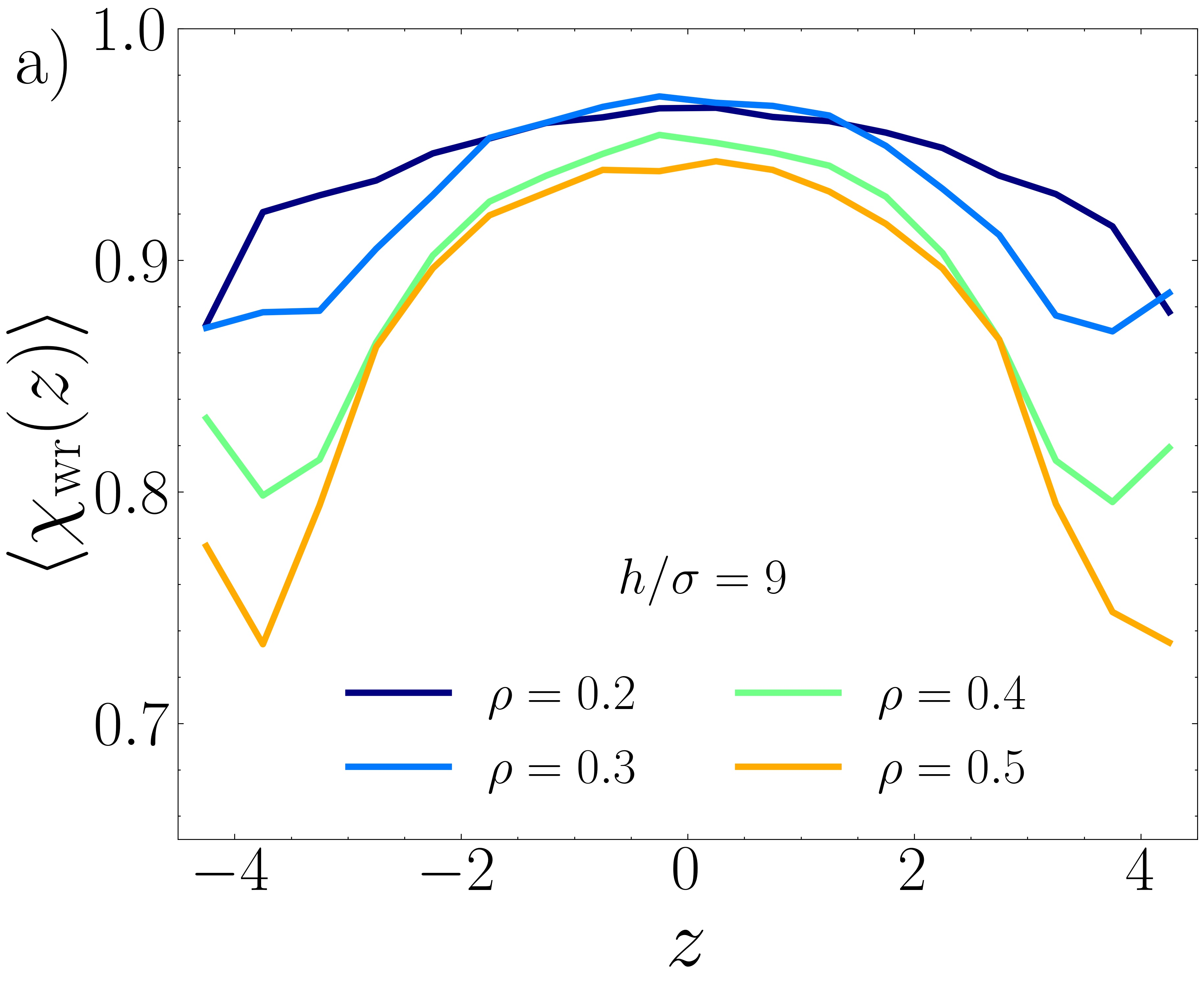}
     \includegraphics[width=0.32\textwidth]{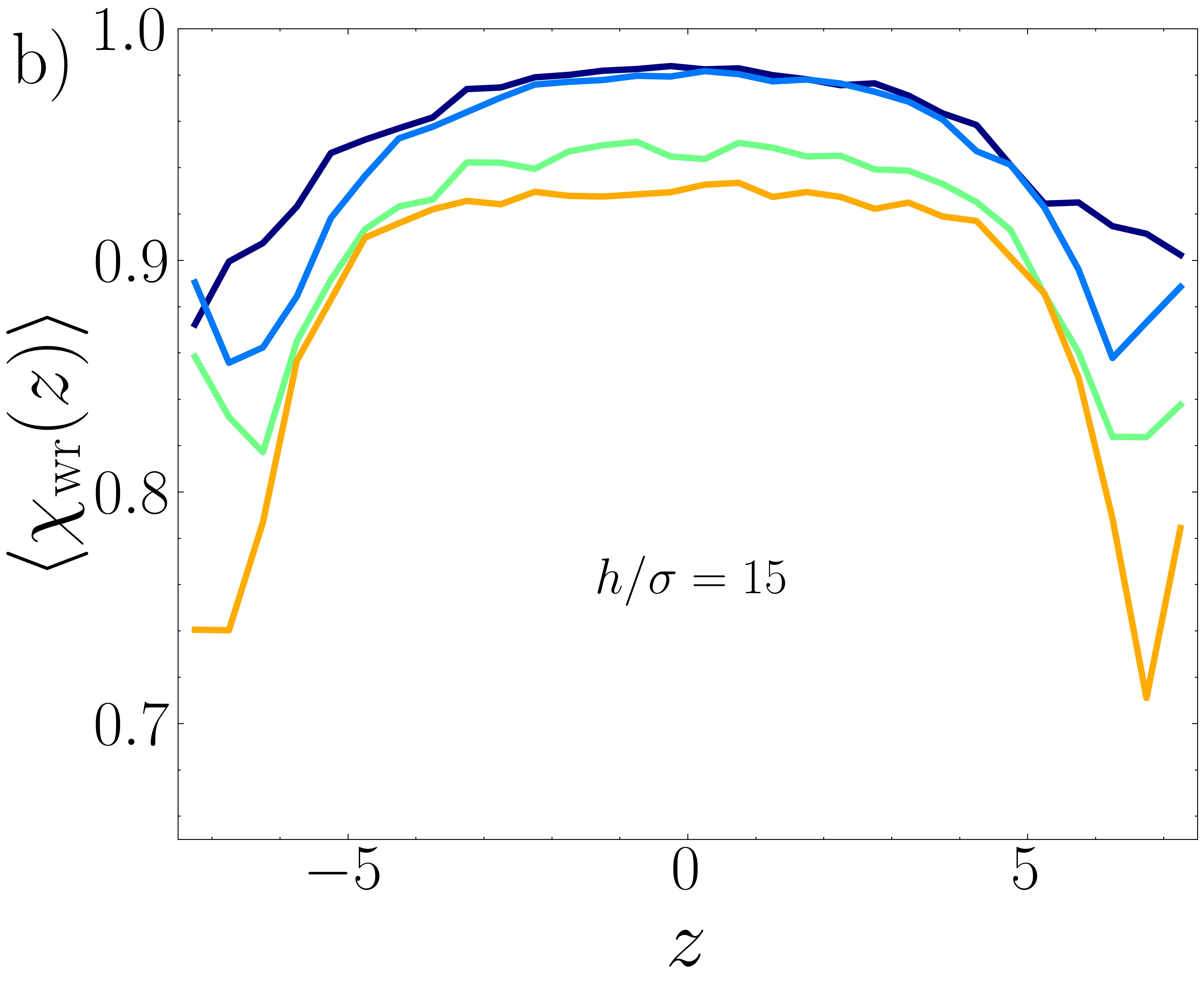}
     \includegraphics[width=0.32\textwidth]{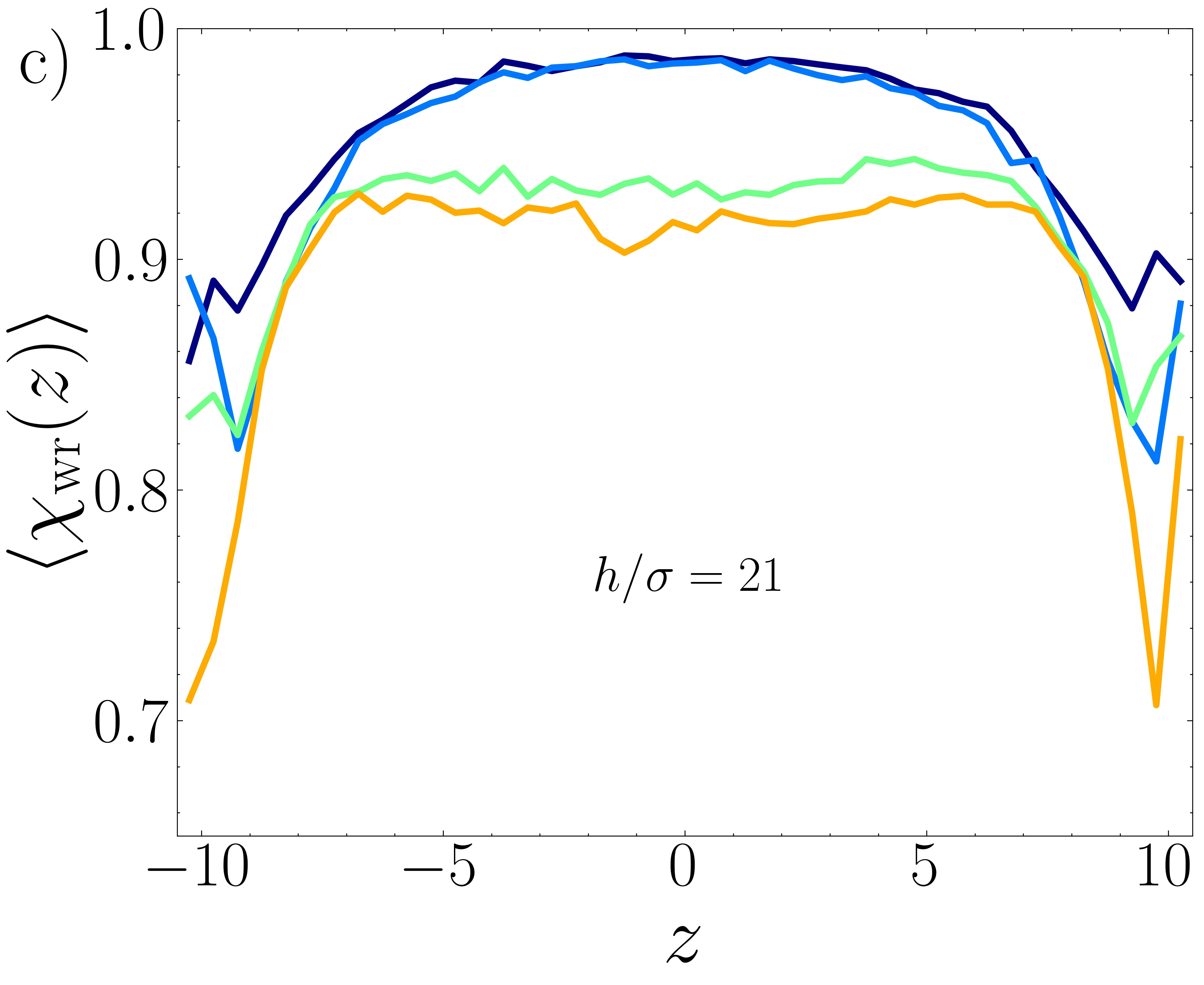}
     \includegraphics[width=0.32\textwidth]{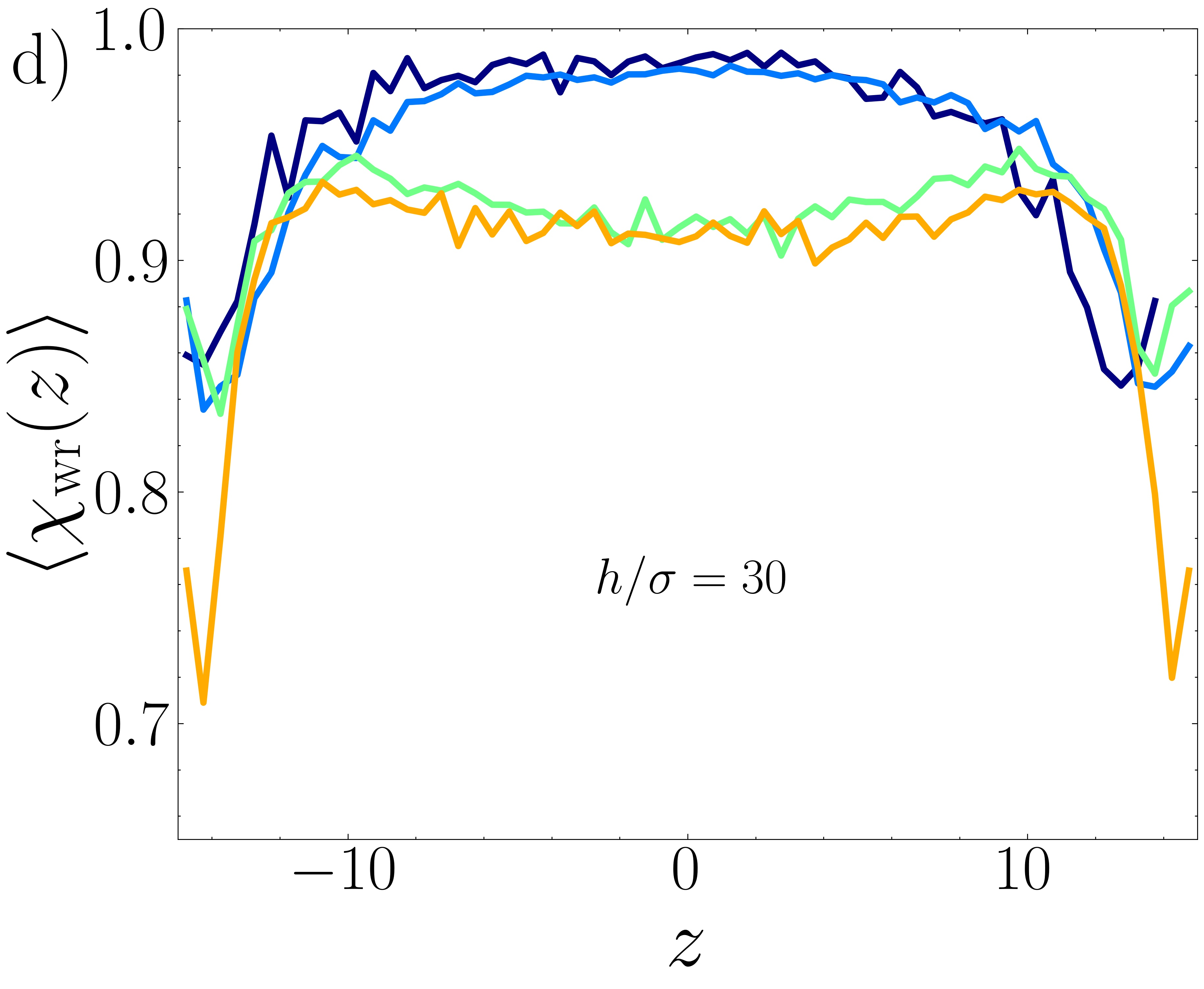}
     \includegraphics[width=0.32\textwidth]{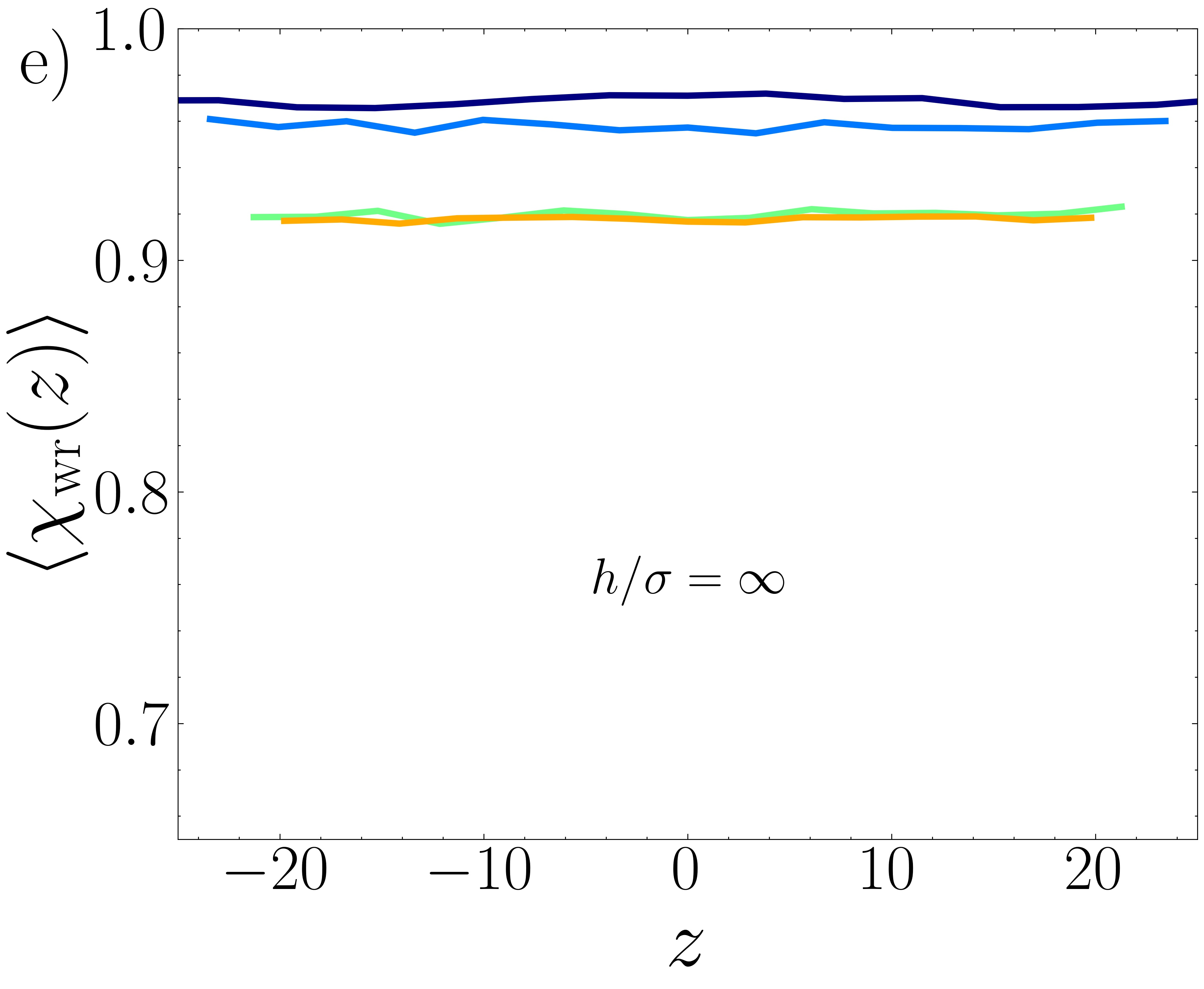}
  \caption{Fraction of significantly ($W>0.5$) wrapped rings as a function of the height $z/h$ or $z/L_z$ in bulk for active rings at different value of the density $\rho$ and (a) $h/\sigma=9$, (b) $h/\sigma=15$, (c) $h/\sigma=21$, (d) $h/\sigma=30$, (a) $h/\sigma=\infty$ (bulk conditions).
  }
\label{fig:chi_suppl}    
\end{figure*}
We report additional results in Fig.~\ref{fig:chi_suppl}, where we report $\chi_{\mathrm{wr}}$ as a function of $z/h$ in confinement or $z/L_z$ in bulk. In each panel we fix the value of $h/\sigma$ and report data at different values of $\rho$. We observe that the effect of the confining walls is more pronounced at high density, irrespective of the confinement: the distribution drops more and more sharply close to the walls with increasing $\rho$. In general, the wrapping decreases with increasing $\rho$, as observed also in the main text as it becomes increasingly difficult for rings to maintain a disk-like conformation and, as such, it becomes more difficult to develop a significant wrapping.

\clearpage
\section{Additional results on the clustering method}

\subsection{Cutoff validation of the alignment-based clustering method}
We report here a validation of the alignment-based clustering, based on the sweep of the cutoff parameter determining the alignment condition. As performed in Ref.~\cite{Miranda2023}, we change one parameter (in this case, the only parameter) and we look for the value that yields a maximum in the number of clusters with the idea that we are not selective enough below this value and we are too strict above this value. Alternatively, we can look for the point of maximum slope in the cluster fraction.\\The only free parameter in the alignment-based clustering is the alignment condition, ruled by $\delta_{\mathrm{cut}}$: two neighboring rings are aligned if the absolute value of the dot product between their directors is larger than $\delta_{\mathrm{cut}}$. Ultimately, we choose a common value for all the systems as, qualitatively, our results will be weakly dependent on the precise choice of this cutoff value.\\  
\begin{figure*}[h]
    \centering
    \includegraphics[width=0.35\textwidth]{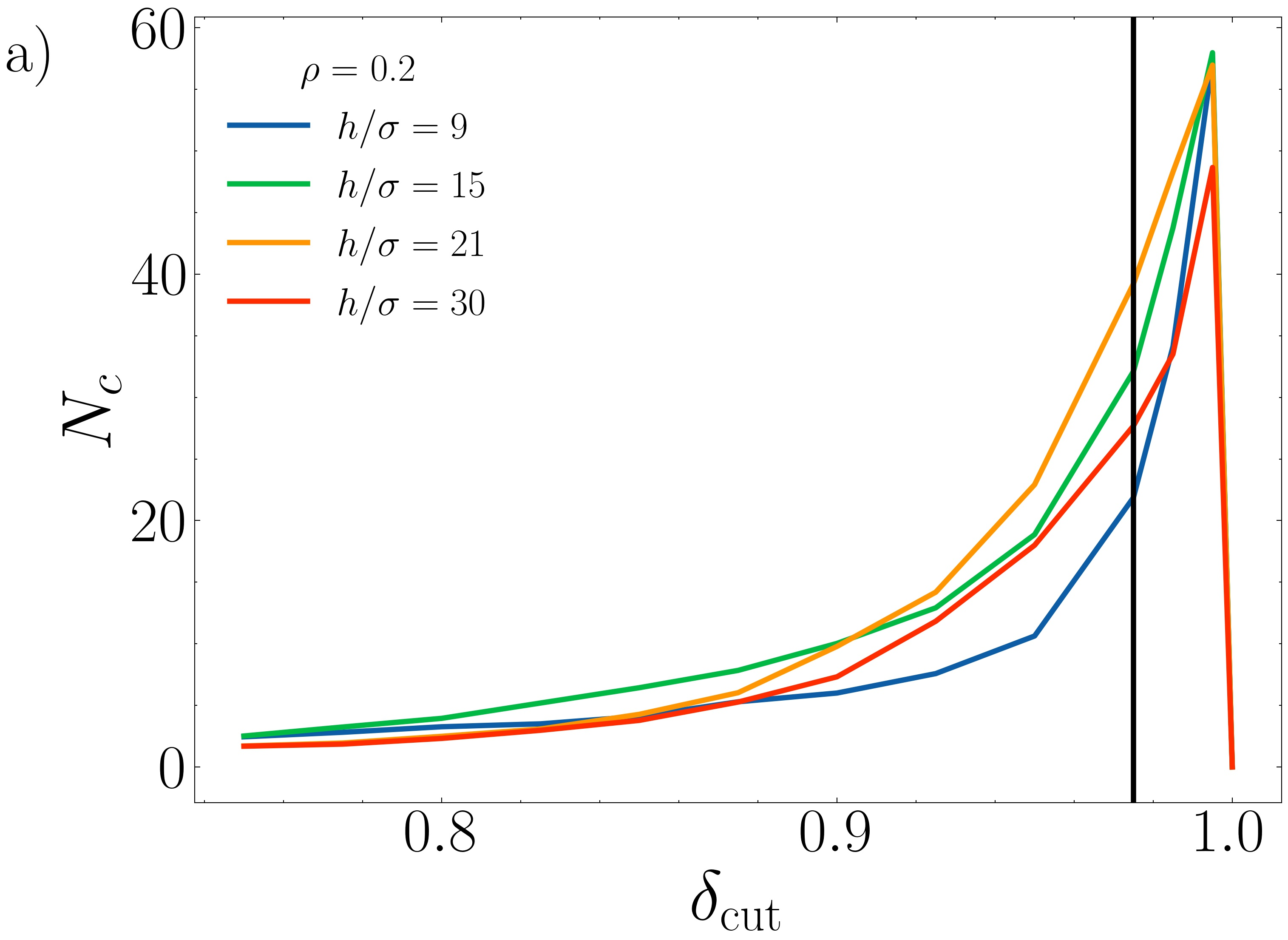}
    \includegraphics[width=0.35\textwidth]{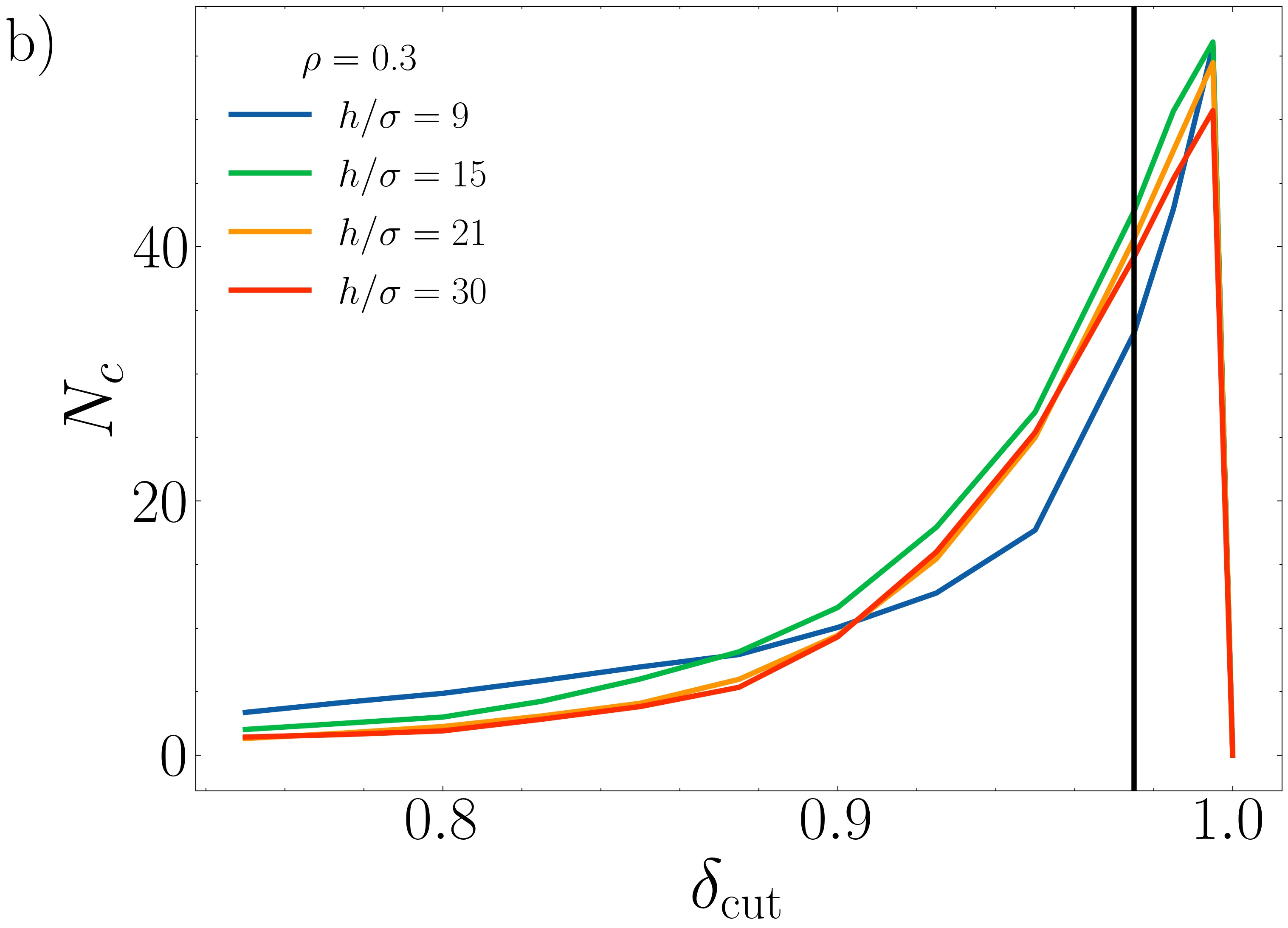}
    \includegraphics[width=0.35\textwidth]{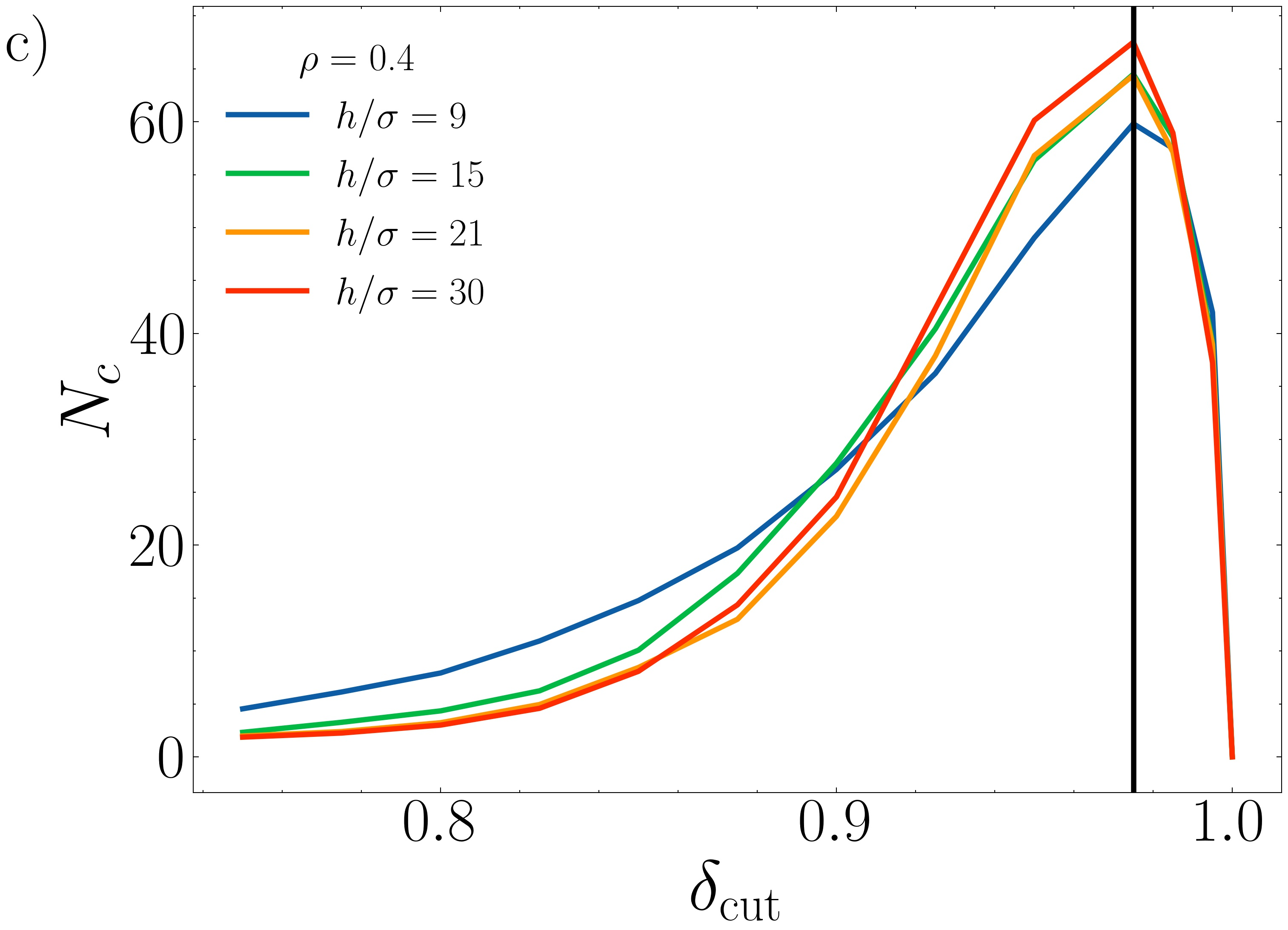}
    \includegraphics[width=0.35\textwidth]{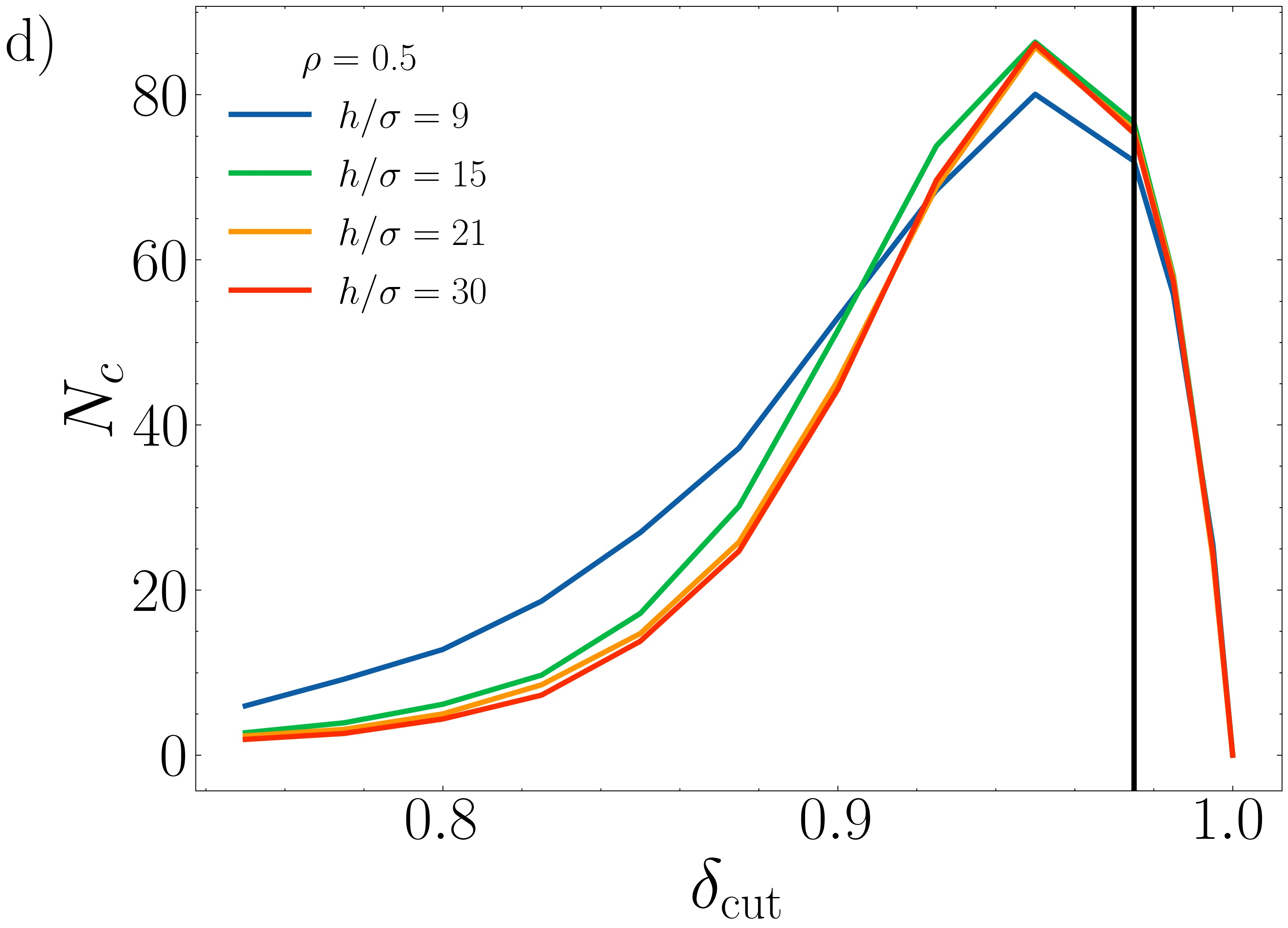}
    \caption{Number of clusters $N_c$ for the alignment-based clustering method as a function of the cutoff parameter $\delta_{\mathrm{cut}}$ for different values $h/\sigma$ (reported with different colors) and a) $\rho =0.2$, b) $\rho=0.3$, c) $\rho=0.4$, d) $\rho=0.5$. The black vertical line marks the chosen cutoff value $\delta_{\mathrm{cut}}=0.975$. }
    \label{fig:Nc_deltacut}
\end{figure*}
We report the number of clusters $N_c$ as a function of the cutoff parameter $\delta_{\mathrm{cut}}$ for different values $h/\sigma$ and $\rho$ in Fig.~\ref{fig:Nc_deltacut}. We notice that the most probable value depends on $\rho$, decreasing with increasing $\rho$, and is independent on $h/\sigma$: with increasing density the crowding increases and it becomes more difficult for rings to align. As such, having more clusters at a lower cutoff should be expected. However, as noticed also in the main text, the absolute value of the number of clusters increases with increasing $\rho$.\\
\begin{figure*}[t]
    \centering
    \includegraphics[width=0.35\textwidth]{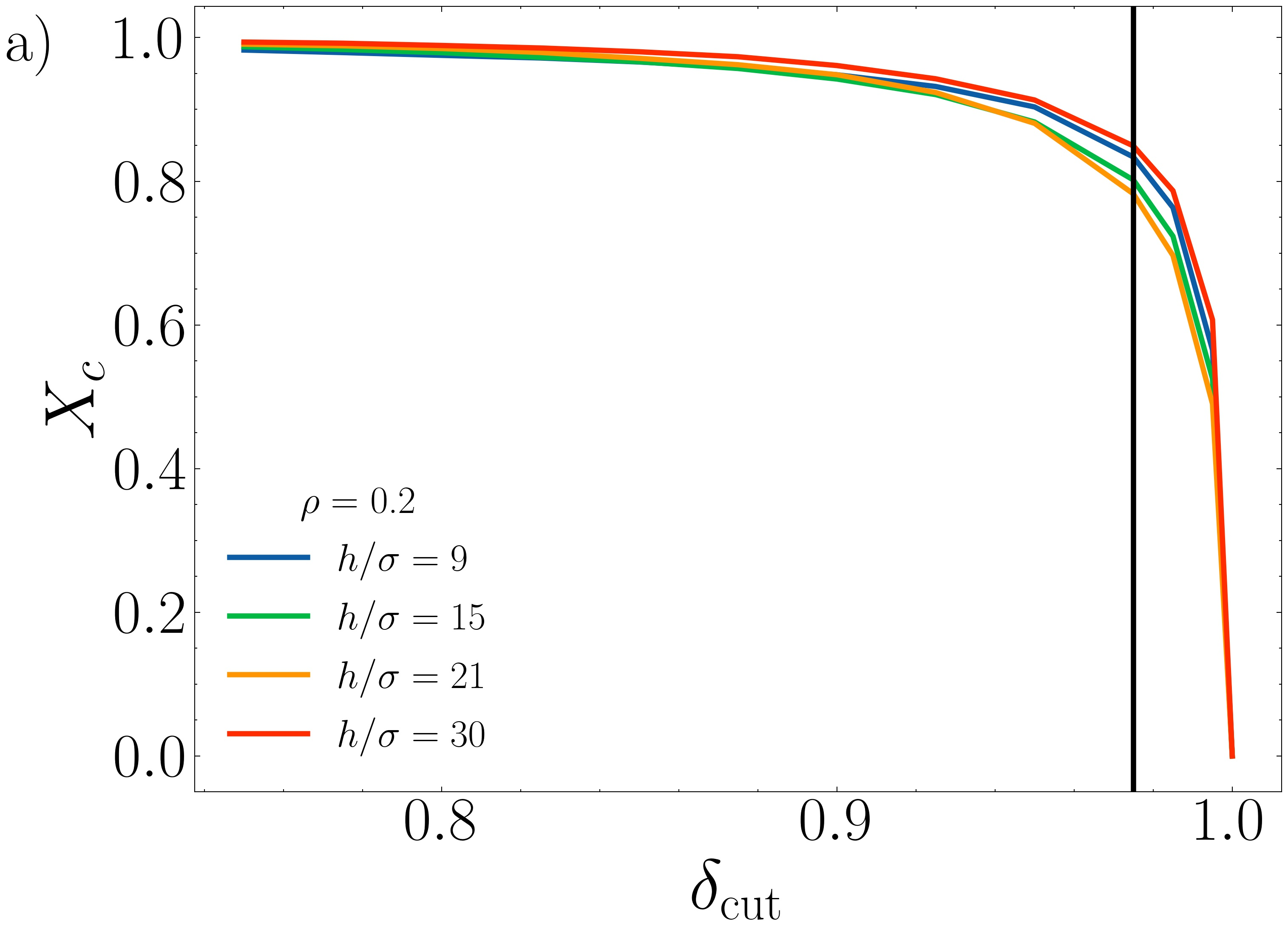}
    \includegraphics[width=0.35\textwidth]{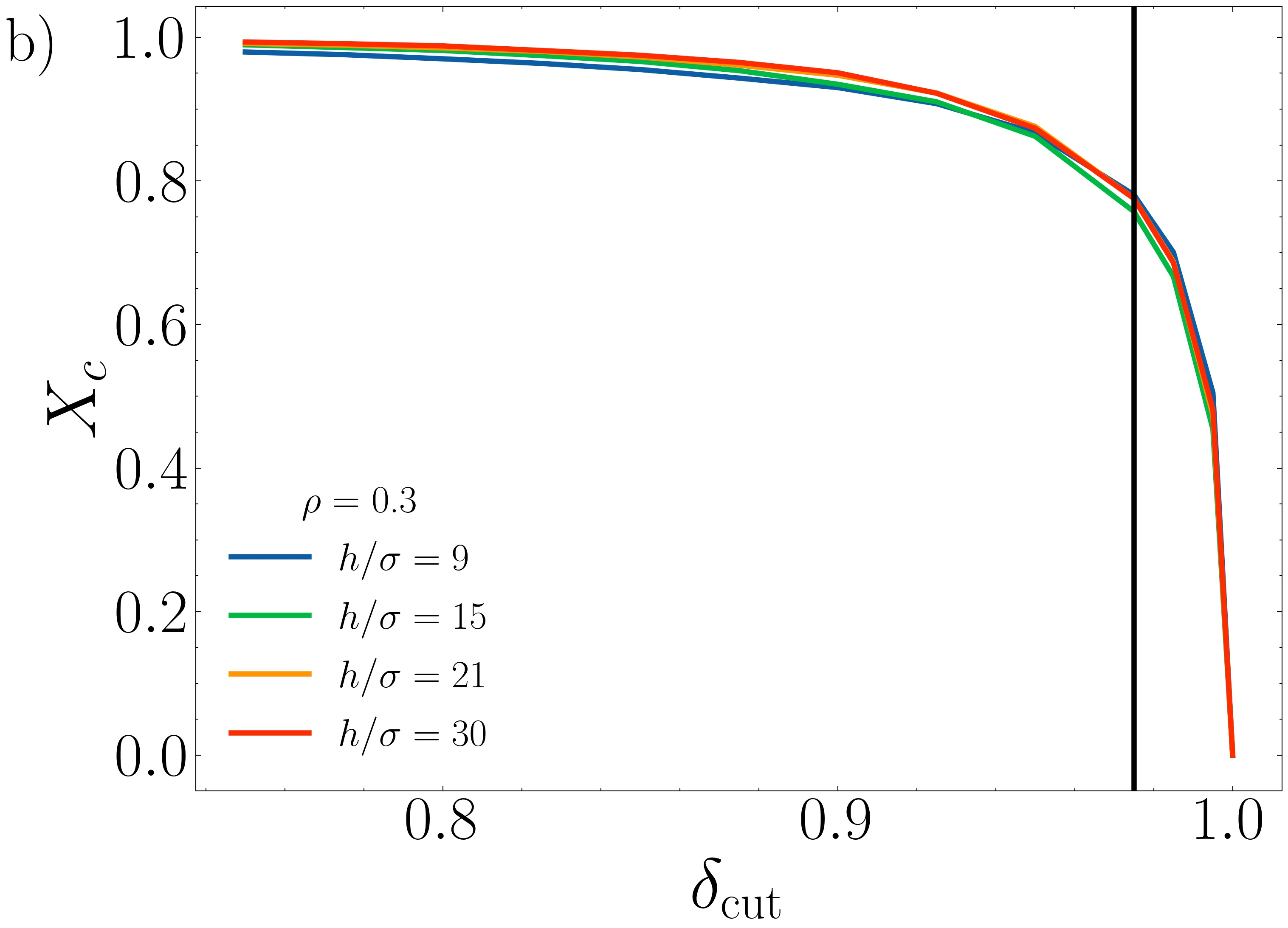}
    \includegraphics[width=0.35\textwidth]{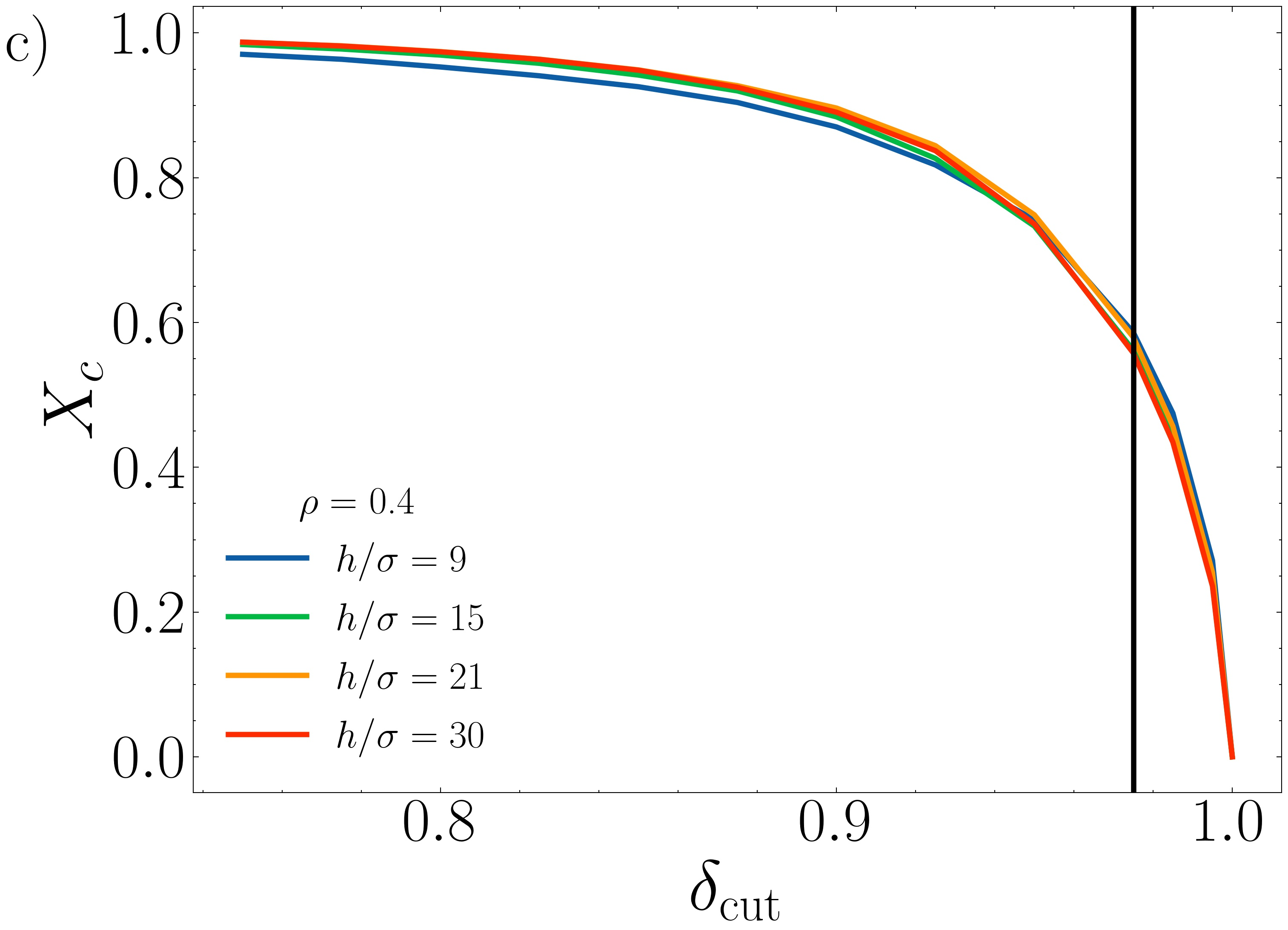}
    \includegraphics[width=0.35\textwidth]{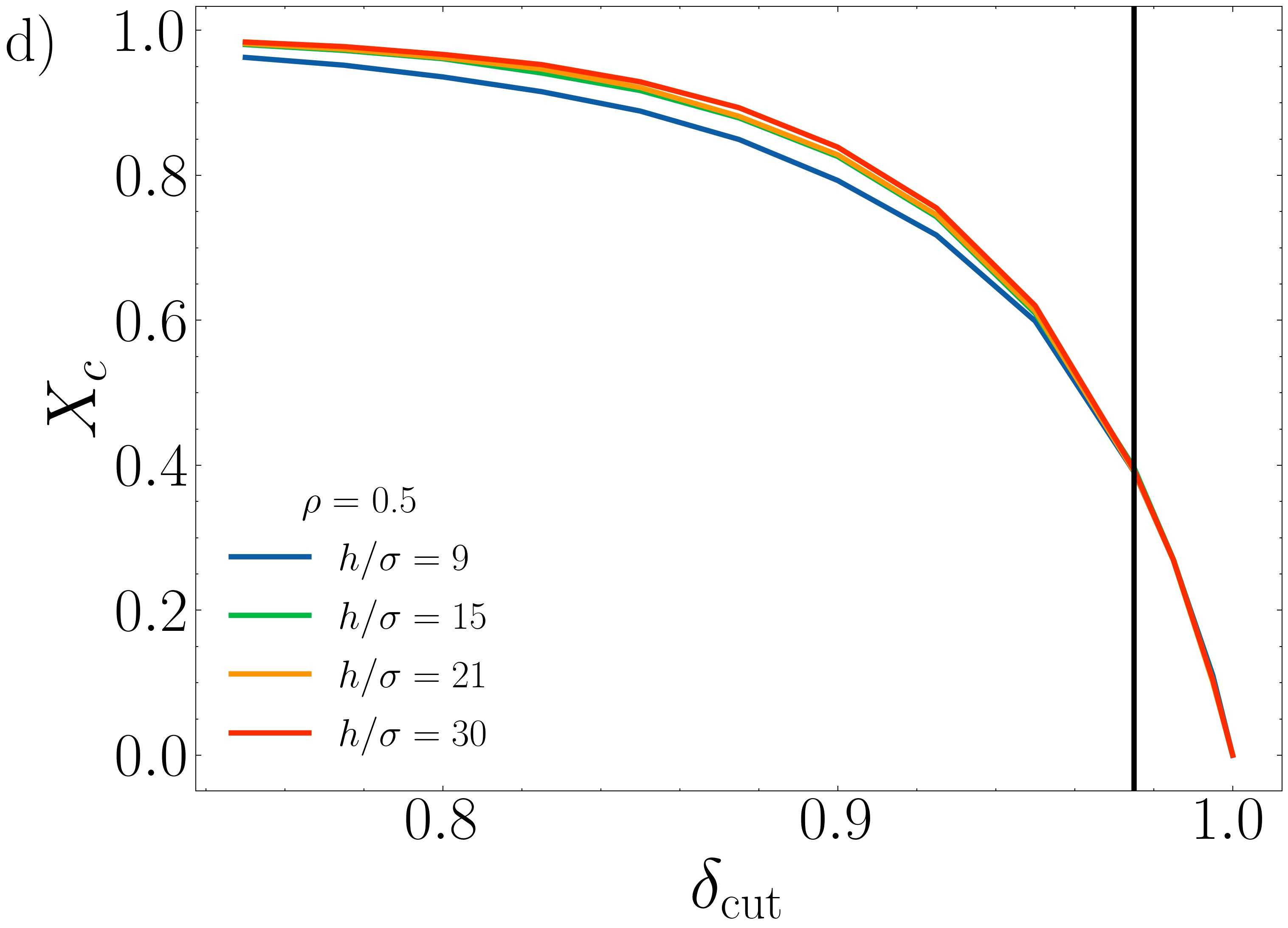}
    \caption{Cluster fraction $X_c$ for the alignment-based clustering method as a function of the cutoff parameter $\delta_{\mathrm{cut}}$ for different values $h/\sigma$ (reported with different colors) and a) $\rho =0.2$, b) $\rho=0.3$, c) $\rho=0.4$, d) $\rho=0.5$. The black vertical line marks the chosen cutoff value $\delta_{\mathrm{cut}}=0.975$. }
    \label{fig:Xc_deltacut}
\end{figure*}
We report the cluster fraction $X_c$ as a function of the cutoff parameter $\delta_{\mathrm{cut}}$ for different values $h/\sigma$ and $\rho$ in Fig.~\ref{fig:Xc_deltacut}.  We notice that the cluster fraction drops to zero much more rapidly at low density than at high density: this indicates that a tight alignment condition is more important at low density than at high density. All in all, we choose a common value for all the systems, $\delta_{\mathrm{cut}}=0.975$, marked in all panels of Figs.~\ref{fig:Nc_deltacut},~\ref{fig:Xc_deltacut} with a black vertical line. We do this for practical reasons: the exact value would be relatively close and, as already said, our results is qualitatively robust with respect to the precise choice of $\delta_{\mathrm{cut}}=0.975$.

\clearpage
\subsection{Cluster size distribution}
We report here additional information on the characterization of the clusters by plotting the cluster size distribution, that is, the probability density function of having a cluster of a certain size. We compute the clusters as detailed in the main text, using the alignment based criterion on Voronoi neighbors.\\
\begin{figure*}[h]
    \centering
    \includegraphics[width=0.33\textwidth]{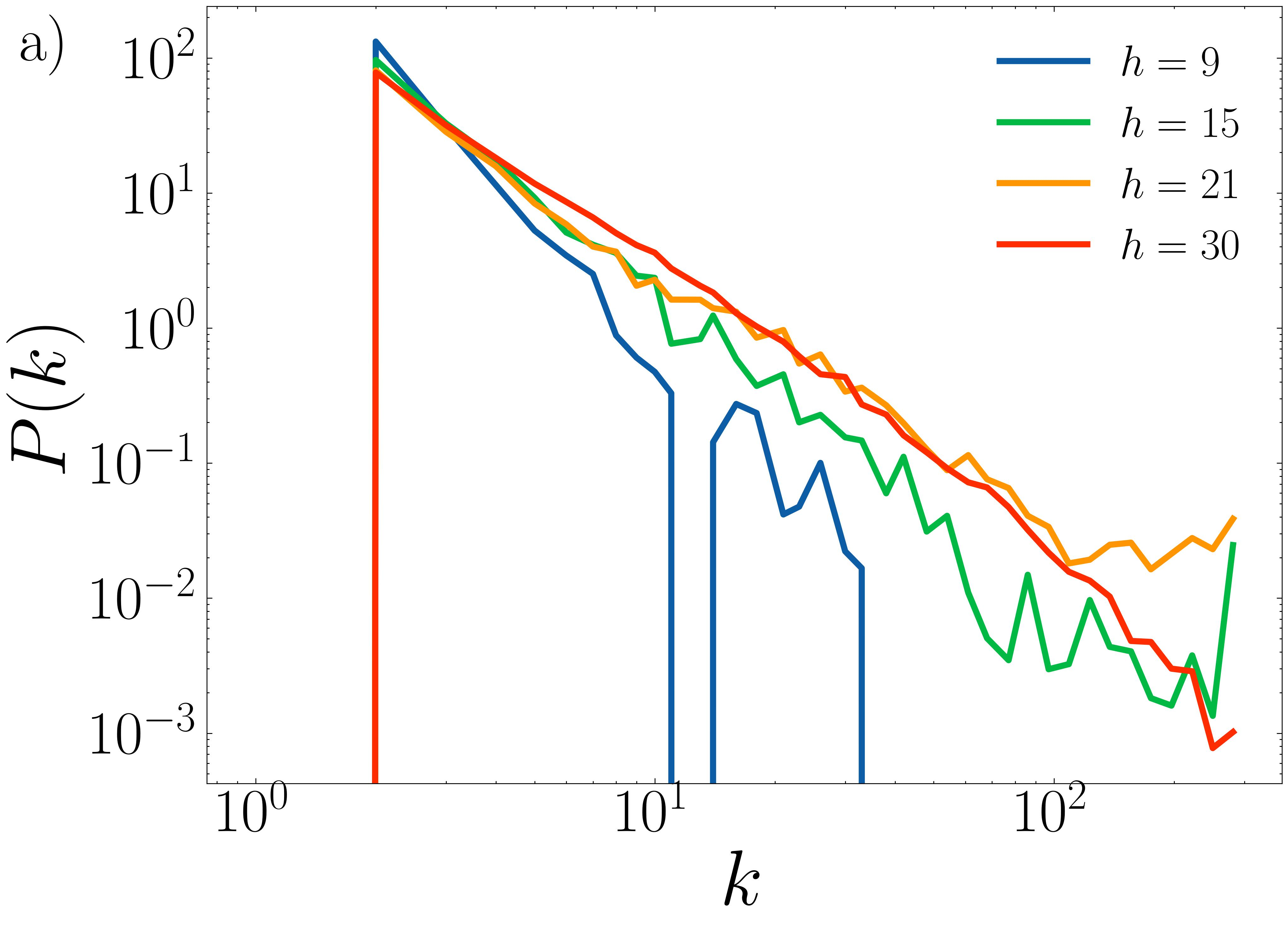}
    \includegraphics[width=0.33\textwidth]{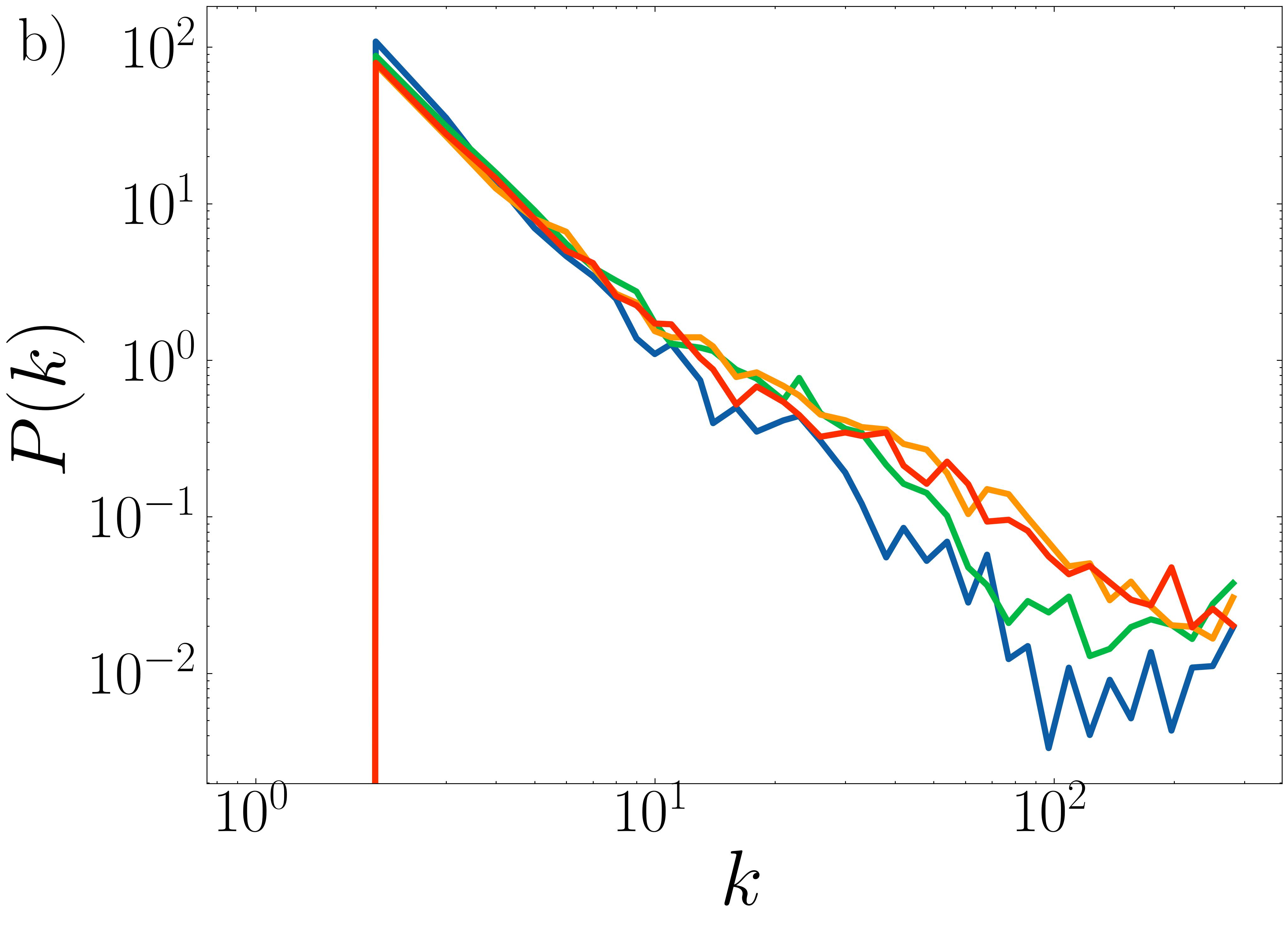}
    \includegraphics[width=0.33\textwidth]{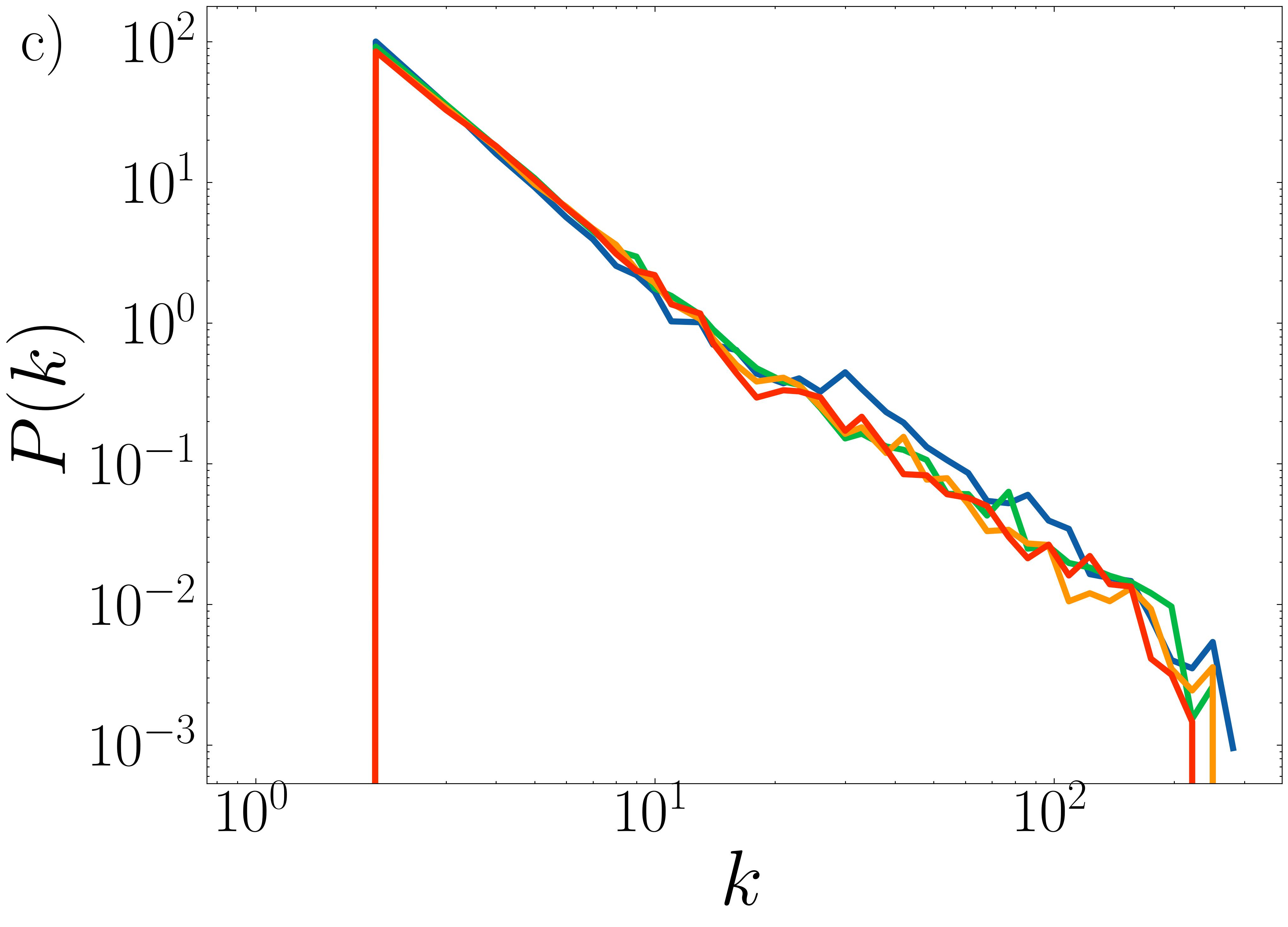}
    \includegraphics[width=0.33\textwidth]{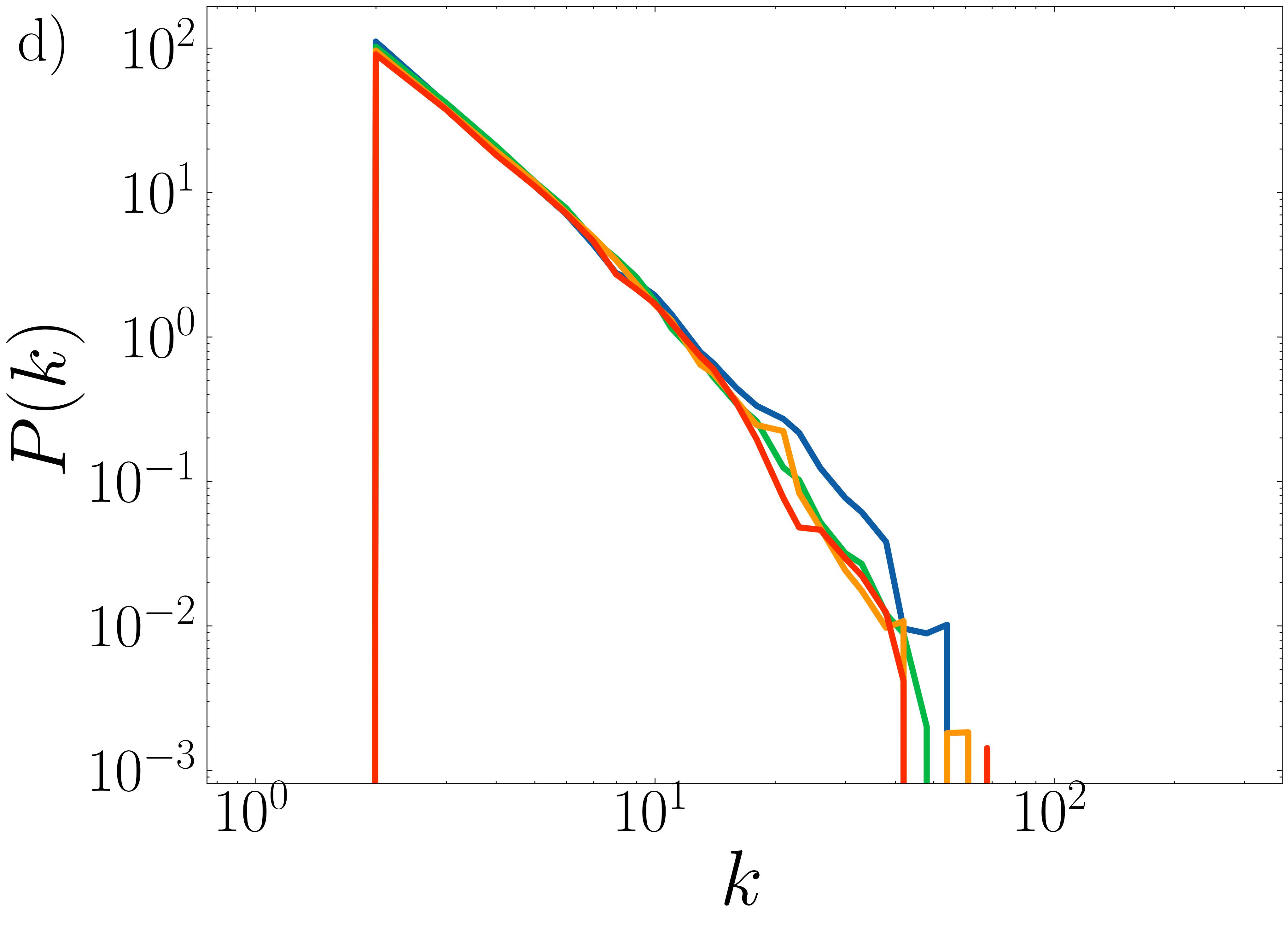}
    \caption{Cluster size distribution $P(k)$ for the alignment-based clustering method as a function of the cluster size $k$ for different values of $h/\sigma$ (reported with different colors) and a) $\rho=0.2$, b) $\rho=0.3$, c) $\rho=0.4$, d) $\rho=0.5$.}
    \label{fig:clustersizedistro}
\end{figure*}
We report our results for the cluster size distribution $P(k)$ as a function of the cluster size $k$ for different values of $h/\sigma$ and $\rho$ in Fig.~\ref{fig:clustersizedistro}. Notice that the minimum value and the most probable one in all cases is $k=2$, that is, a cluster made of two rings. In all cases, except $\rho=0.2$, the cluster size distributions are independent of $h/\sigma$; at $\rho=0.2$ the probability of finding very large clusters decreases with decreasing $h/\sigma$. Further, for all cases but $\rho=0.5$ the cluster size distributions are characterized by a power-law scaling; at $\rho=0.5$, very large clusters are much less probable.   


\clearpage
\section{Interplay between alignment and entanglement - additional results}\label{sec:wrappalign_supl}

\begin{figure}[h!]
    \centering
    \includegraphics[width=0.45\textwidth]{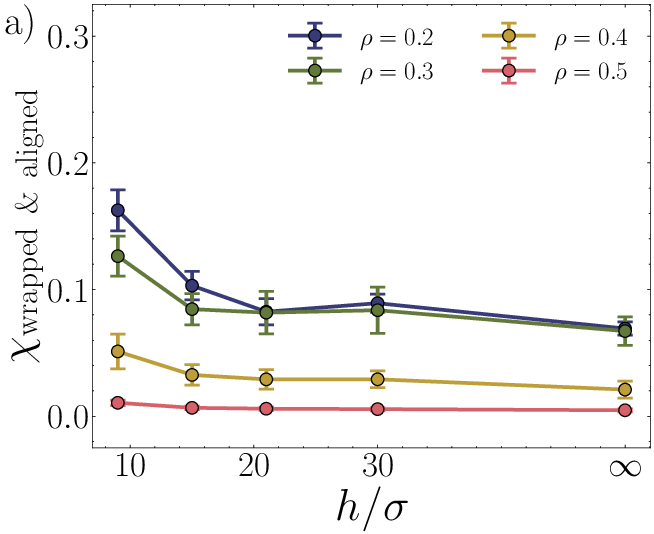}
    \includegraphics[width=0.45\textwidth]{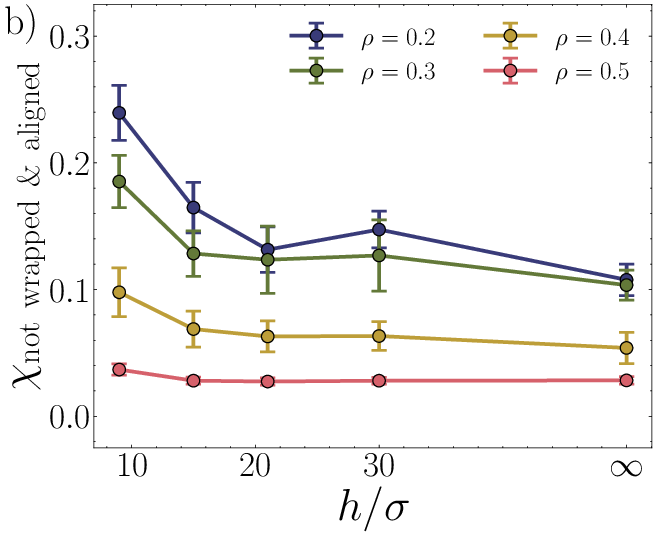}
    \caption{Fraction of {significantly} wrapped Voronoi neighboring pairs, whose directors are a) aligned or b) not aligned as a function of $h/\sigma$ for different values of the density. The alignment criterion is the same one used for clustering, $ | \vec{d}_i \cdot \vec{d}_j | \geq \delta_{\mathrm{cut}} = 0.975$, {and significantly wrapped rings have $W_{\mathrm{max}} \geq 0.5$}. 
    }
    \label{fig:wrapped_aligned}
\end{figure}

\begin{figure}[h]
    \centering
    \includegraphics[width=0.45\textwidth]{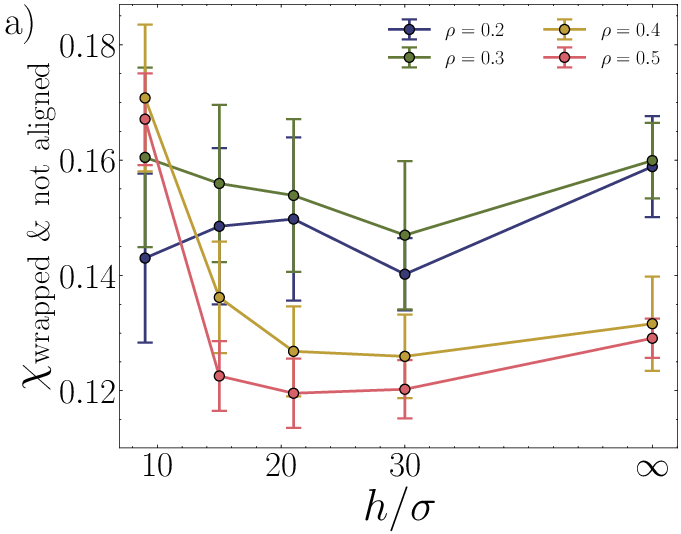}
    \includegraphics[width=0.45\textwidth]{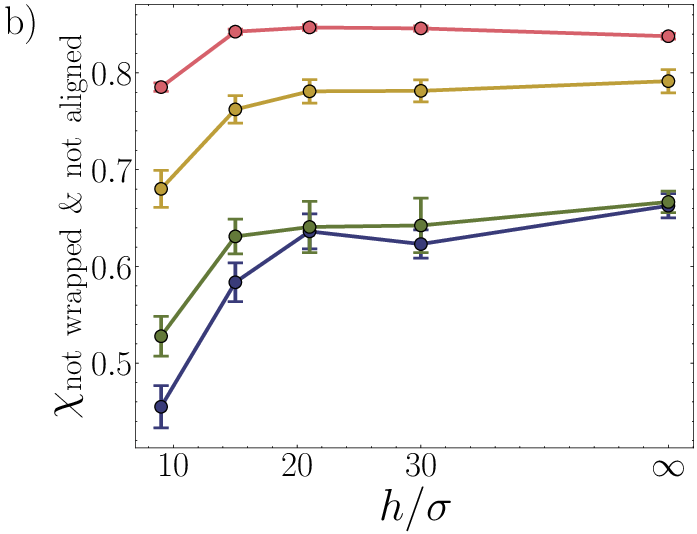}
    \caption{Fraction of neighbouring active rings that are a) {Significantly} entangled but not aligned or b) neither not {significantly} entangled nor aligned as a function of $\rho$ for different values of $h/\sigma$.
    }
    \label{fig:wrappedaligned_suppl}
\end{figure}

We complement the data reported on 
the main text, categorizing how the neighboring active rings pairs split into four different categories: (i) aligned and significantly entangled, (ii) aligned but not significantly entangled, (iii) not aligned but significantly entangled, (iv) not aligned and not significantly entangled and how the different fractions depend on the degree of confinement and on the density. 
As in the main text, neighboring rings are identified via a Voronoi tessellation and we classify these pairs as ``significantly entangled'' {if $W_{\mathrm{max}} > 0.5$ (see Sec.~\ref{sec:wrap_def})} and as ``aligned'' if $ | \vec{d}_i \cdot \vec{d}_j | \geq 0.975$,  which is the criterion used for {identifying} the clusters. 
%
%
In Fig.~\ref{fig:wrapped_aligned}, we plot the fraction of neighboring {aligned} rings
that are {significantly entangled} (Fig.~\ref{fig:wrapped_aligned} a) or {not significantly entangled} (Fig.~\ref{fig:wrapped_aligned} b) as a function of confinement, $h/\sigma$, for various densities, $\rho$. {Notice that the majority of neighbouring pairs are not aligned; their fraction is reported in Fig.~\ref{fig:wrappedaligned_suppl}.} 

Notably, the current definition of significant entanglement does not match the sweet spot emerging from the joint probability distribution; as such, the relevant information is split between the two panels. However, it is interesting to notice that, with this agnostic definition, we are able to still recover the correct trends,  as highlighted in the main text. Further, notice that an overall fraction of $\sim 15$\% of aligned rings are still able to produce a significant level of self-organization at $\rho=0.4$. 

We report the data for non-aligned rings in Fig.~\ref{fig:wrappedaligned_suppl}. As visible in Fig.~\ref{fig:wrappedaligned_suppl}a), the fraction of rings that are significantly entangled but not aligned is quite small and similar for all values of $\rho$. Conversely, the fraction of rings that are not aligned and not wrapped is the largest fraction practically for every values of $\rho$ and $h/\sigma$. Roughly speaking, this category gathers contributions from rings that are not in clusters as well as from rings that belong to different clusters. 

\subsection{Z-score of passive systems}

We report here the Z-score measured in passive systems. The Z-score is defined in the main text: briefly, we compute the Mutual Information (MI) between wrapping and alignment of Voronoi neighbors (Eq.~(8) of the main text) and we then compare it with the MI measured from completely random samples to compute the Z-score (Eq.~(9) of the main text).
\begin{figure}
    \centering
    \includegraphics[width=0.45\textwidth]{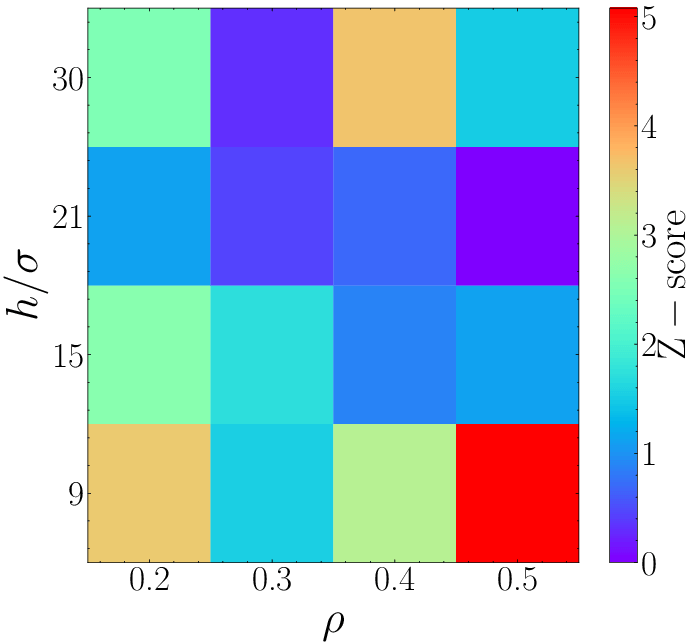}
    \caption{Color plot of the Z-score test between alignment and wrapping of Voronoi neighbors in the $\rho$-$h/\sigma$ plane for systems of passive rings ($\mathrm{Pe}=0$). }
    \label{fig:zscore_suppl}
\end{figure}
We report the results in Fig.~\ref{fig:zscore_suppl}. At variance with the active case (Fig.~8 of the main text), we do not notice any pattern in the results; in addition, the Z-score is low (of order unity), showing that in passive systems the MI between wrapping and alignment of neighboring rings is compatible with the MI of a purely random sample.

\bibliography{bibliography}